\documentclass[twocolumn]{aastex61}

\usepackage{amsmath}

\shorttitle{Ly$\alpha$ Blobs at $z=4.9-7.0$}
\shortauthors{Zhang et al.}

\begin{document}


\title{CHORUS. III. Photometric and Spectroscopic Properties of \\
L\lowercase{y}$\alpha$ Blobs at $\lowercase{z}=4.9-7.0$}

\author{Haibin Zhang} 
\affil{\rm Institute for Cosmic Ray Research, The University of Tokyo, 5-1-5 Kashiwanoha, Kashiwa, Chiba 277-8582, Japan; \url{hbz@icrr.u-tokyo.ac.jp}}
\affil{\rm Department of Physics, Graduate School of Science, The University of Tokyo, 7-3-1 Hongo, Bunkyo-ku, Tokyo 113-0033, Japan}

\author{Masami Ouchi}
\affil{\rm Institute for Cosmic Ray Research, The University of Tokyo, 5-1-5 Kashiwanoha, Kashiwa, Chiba 277-8582, Japan; \url{hbz@icrr.u-tokyo.ac.jp}}
\affil{\rm Kavli Institute for the Physics and Mathematics of the Universe (WPI), The University of Tokyo, Kashiwa 277-8583, Japan}

\author{Ryohei Itoh}
\affil{\rm Institute for Cosmic Ray Research, The University of Tokyo, 5-1-5 Kashiwanoha, Kashiwa, Chiba 277-8582, Japan; \url{hbz@icrr.u-tokyo.ac.jp}}
\affil{\rm Department of Physics, Graduate School of Science, The University of Tokyo, 7-3-1 Hongo, Bunkyo-ku, Tokyo 113-0033, Japan}

\author{Takatoshi Shibuya}
\affil{\rm Kitami Institute of Technology, 165 Koen-cho, Kitami, Hokkaido 090-8507, Japan}

\author{Yoshiaki Ono}
\affil{\rm Institute for Cosmic Ray Research, The University of Tokyo, 5-1-5 Kashiwanoha, Kashiwa, Chiba 277-8582, Japan; \url{hbz@icrr.u-tokyo.ac.jp}}

\author{Yuichi Harikane}
\affil{\rm National Astronomical Observatory of Japan, 2-21-1 Osawa, Mitaka, Tokyo 181-8588, Japan}

\author{Akio K. Inoue}
\affil{\rm Department of Environmental Science and Engineering, Faculty of Design Technology, Osaka Sangyo University, 3-1-1 Nakagaito, Daito, Osaka 574-8530, Japan}
\affil{\rm Department of Physics, School of Advanced Science and Engineering, Waseda University, 3-4-1 Okubo, Shinkuju, Tokyo 169-8555, Japan}
\affil{\rm Waseda Research Institute for Science and Engineering, 3-4-1 Okubo, Shinjuku, Tokyo 169-8555, Japan}

\author{Michael Rauch}
\affil{\rm The Observatories of the Carnegie Institution for Science, 813 Santa Barbara Street, Pasadena, CA 91101, USA}

\author{Shotaro Kikuchihara}
\affil{\rm Institute for Cosmic Ray Research, The University of Tokyo, 5-1-5 Kashiwanoha, Kashiwa, Chiba 277-8582, Japan; \url{hbz@icrr.u-tokyo.ac.jp}}
\affil{\rm Department of Astronomy, Graduate School of Science, The University of Tokyo, 7-3-1 Hongo, Bunkyo-ku, Tokyo 113-0033, Japan}

\author{Kimihiko Nakajima}
\affil{\rm National Astronomical Observatory of Japan, 2-21-1 Osawa, Mitaka, Tokyo 181-8588, Japan}

\author{Hidenobu Yajima}
\affil{\rm Center for Computational Sciences, University of Tsukuba, Ten-nodai, 1-1-1 Tsukuba, Ibaraki 305-8577, Japan}

\author{Shohei Arata}
\affil{\rm Department of Earth and Space Science, Graduate School of Science, Osaka University, Toyonaka, Osaka 560-0043, Japan}


\author{Makito Abe}
\affil{\rm Center for Computational Sciences, University of Tsukuba, Ten-nodai, 1-1-1 Tsukuba, Ibaraki 305-8577, Japan}

\author{Ikuru Iwata}
\affil{\rm National Astronomical Observatory of Japan, 2-21-1 Osawa, Mitaka, Tokyo 181-8588, Japan}
\affil{\rm Department of Astronomical Science, SOKENDAI (The Graduate University for Advanced
Studies), 2-21-1 Osawa, Mitaka, Tokyo 181-8588, Japan}

\author{Nobunari Kashikawa}
\affil{\rm National Astronomical Observatory of Japan, 2-21-1 Osawa, Mitaka, Tokyo 181-8588, Japan}
\affil{\rm Department of Astronomy, Graduate School of Science, The University of Tokyo, 7-3-1 Hongo, Bunkyo-ku, Tokyo 113-0033, Japan}

\author{Satoshi Kawanomoto}
\affil{\rm National Astronomical Observatory of Japan, 2-21-1 Osawa, Mitaka, Tokyo 181-8588, Japan}

\author{Satoshi Kikuta}
\affil{\rm National Astronomical Observatory of Japan, 2-21-1 Osawa, Mitaka, Tokyo 181-8588, Japan}
\affil{\rm Department of Astronomical Science, SOKENDAI (The Graduate University for Advanced
Studies), 2-21-1 Osawa, Mitaka, Tokyo 181-8588, Japan}

\author{Masakazu Kobayashi}
\affil{\rm Faculty of Natural Sciences, National Institute of Technology, Kure College, 2-2-11
Agaminami, Kure, Hiroshima 737-8506, Japan}

\author{Haruka Kusakabe}
\affil{\rm Observatoire de Gen\`{e}ve, Universit\'e de Gen\`{e}ve, 51 Ch. des Maillettes, 1290 Versoix, Switzerland}

\author{Ken Mawatari}
\affil{\rm Institute for Cosmic Ray Research, The University of Tokyo, 5-1-5 Kashiwanoha, Kashiwa, Chiba 277-8582, Japan; \url{hbz@icrr.u-tokyo.ac.jp}}

\author{Tohru Nagao}
\affil{\rm Research Center for Space and Cosmic Evolution, Ehime University, 2-5 Bunkyo-cho, Matsuyama, Ehime 790-8577, Japan}

\author{Kazuhiro Shimasaku}
\affil{\rm Department of Astronomy, Graduate School of Science, The University of Tokyo, 7-3-1 Hongo, Bunkyo-ku, Tokyo 113-0033, Japan}
\affil{\rm Research Center for the Early Universe, Graduate School of Science, The University of Tokyo, 7-3-1 Hongo, Bunkyo, Tokyo 113-0033, Japan}

\author{Yoshiaki Taniguchi}
\affil{\rm The Open University of Japan, Wakaba 2-11, Mihama-ku, Chiba 261-8586, Japan}



\begin{abstract}

We report the Subaru Hyper Suprime-Cam (HSC) discovery of two Ly$\alpha$ blobs (LABs), dubbed 
z70-1 and z49-1 at $z=6.965$ and $z=4.888$ respectively, that are Ly$\alpha$ emitters with a bright ($\log L_{\rm Ly\alpha}/{\rm [erg\ s^{-1}]}>43.4$) and spatially-extended Ly$\alpha$ emission, and present the photometric and spectroscopic properties of a total of seven LABs; the two new LABs and five previously-known LABs at $z=5.7-6.6$. The z70-1 LAB shows the extended Ly$\alpha$ emission with a scale length of $1.4\pm 0.2$ kpc, about three times larger than the UV continuum emission, making z70-1 the most distant LAB identified to date. All of the 7 LABs, except z49-1, exhibit no AGN signatures such as X-ray emission, {\sc Nv}$\lambda$1240 emission, or Ly$\alpha$ line broadening, while z49-1 has a strong {\sc Civ}$\lambda$1548 emission line indicating an AGN on the basis of the UV-line ratio diagnostics. We carefully model the point-spread functions of the HSC images, and conduct two-component exponential profile fitting to the extended Ly$\alpha$ emission of the LABs. The Ly$\alpha$ scale lengths of the core (star-forming region) and the halo components are $r_{\rm c}=0.6-1.2$ kpc and $r_{\rm h}=2.0-13.8$ kpc, respectively. 
The average $r_{\rm h}$ of the LABs falls on the extrapolation of the $r_{\rm h}$-Ly$\alpha$ luminosity relation of the Ly$\alpha$ halos around VLT/MUSE star-forming galaxies at the similar redshifts, suggesting that typical LABs at $z\gtrsim5$ are not special objects, but star-forming galaxies at the bright end.

\end{abstract}

\keywords{galaxies: formation - galaxies: evolution - galaxies: high-redshift - cosmology: observations}



\section{Introduction} \label{sec:intro}

Ly$\alpha$ emitters (LAEs) are important objects to study the formation and evolution of star-forming galaxies at high redshifts where the redshifted Ly$\alpha$ emission becomes observable with ground-based telescopes (e.g. \citealt{ouchi2003, ouchi2008}). Previous narrowband imaging surveys have identified LAEs with very luminous Ly$\alpha$ emission (log $(L_{\rm Ly\alpha}/{\rm [erg\ s^{-1}]})\gtrsim43.4$) and a large isophotal area ($\gtrsim150$ kpc$^2$) at $z\sim2-7$. These luminous and spatially extended LAEs are often referred to as Ly$\alpha$ blobs (LABs; e.g. \citealt{keel1999,steidel2000,francis2001,matsuda2004}). One well-known example of LABs is LAB1 (\citealt{steidel2000}) at $z=3.1$, while the most distant ones are Himiko \citep{ouchi2009} and CR7 \citep{sobral2015} at $z=$6.6. LABs are important objects to study massive galaxies and their circumgalactic medium (CGM) in the early universe. Although LABs have been analyzed individually by many studies, the relation between LABs at different epochs of $z\sim3$ (e.g. LAB1) and $z\gtrsim6$ (e.g. Himiko and CR7) is still unclear.

Diffuse Ly$\alpha$ nebulae called Ly$\alpha$ halos (LAHs) are a common feature around LAEs with 
$\log (L_{\rm Ly\alpha}/{\rm [erg\ s^{-1}]})\sim 42-43$ at $z\sim3-6$, and have been identified individually (e.g. \citealt{rauch2008,wisotzki2016, leclercq2017}) or statistically by stacking analysis (e.g. \citealt{hayashino2004,steidel2011, matsuda2012, feldmeier2013, momose2014, momose2016, wisotzki2018}). 
The typical isophotal area of LAHs is smaller than that of LABs at the similar redshift. However, it should be noted that the isophotal area measurement depends on both surface brightness detection limits and radial profiles. 
At the same detection limit, faint LAHs show smaller isophotal areas than bright ones if the radial profile shapes are the same.
Similarly, because the Ly$\alpha$ luminosities of LAHs are fainter than those of LABs by an order of magnitude, LAHs should have smaller isophotal areas than LABs if identical radial profile shapes are assumed. 
Nevertheless, it is still unclear whether LAHs and LABs have similar shapes of Ly$\alpha$ radial profiles.

\citet{konno2016} and \citet{sobral2018} suggest that active galactic nuclei (AGNs) exist in LAEs brighter than a luminosity limit of log $(L_{\rm Ly\alpha}/{\rm [erg\ s^{-1}]})\gtrsim43.4$ at $z\sim2-3$. Because most LABs exceed this luminosity limit, it is expected that LABs have AGN activities. Previously, AGNs have been identified in some LABs (e.g. LAB2 in \citealt{steidel2000}; \citealt{basuzych2004}), while no evidence of AGNs is found in the other LABs (e.g. LAB1; \citealt{geach2007,matsuda2007}). Statistically, \citet{geach2009} investigate 29 LABs at $z=3.09$ and find that $\sim10-30\%$ of the LABs contain AGNs. To explain these observational results, there are two possibilities. One possibility is that all LABs intrinsically have AGNs, and that some AGNs are obscured or too faint to be identified. Another possibility is that there exist two kinds of LABs with and without AGNs. 

Related to the possible AGN activities in LABs, the extended Ly$\alpha$ emission can be explained by several scenarios listed below.

\begin{enumerate}
\item
Fluorescence. There exists some neutral hydrogen gas in the CGM around a galaxy that is heated by an AGN or star formation. The neutral hydrogen gas is photoionized by the radiation from the galaxy center or UV background. Ly$\alpha$ photons are then emitted during the recombination process (e.g. \citealt{masribas2016}). 
\item
Resonant scattering. Ly$\alpha$ photons escape to the CGM from a galaxy center, and are resonantly scattered by the neutral hydrogen in the CGM. This process causes the galaxy having Ly$\alpha$ emission more extended than the UV continuum (e.g. \citealt{lake2015,masribas2017}).
\item
Gravitational cooling radiation. Some inflow streams exist around a galaxy, and accrete onto the galaxy center. Ly$\alpha$ photons are emitted by collisional excitation of neutral hydrogen in the streams. In this radiation process, the streams release their gravitational potential energy (e.g. \citealt{dekel2009}).
\item
Outflows. Multiple supernova explosions in a galaxy produce hot gas outflows. The outflows drive shocked cooling shells that emit Ly$\alpha$ photons (e.g. \citealt{taniguchi2000, mori2004}). 
\item
Satellite galaxies. A central galaxy is surrounded by multiple satellite galaxies that emit Ly$\alpha$ photons during star formation. In this scenario, a galaxy may exhibit both extended Ly$\alpha$ emission and extended UV continuum (e.g. \citealt{masribas2017}).
\end{enumerate}

Because the different possible scenarios are expected to cause different shapes of Ly$\alpha$ radial profiles, various studies have tried to pinpoint the origin of extended Ly$\alpha$ emission by comparing Ly$\alpha$ radial profiles from models with those from observations (e.g. \citealt{lake2015,masribas2016,masribas2017}). However, these studies target LAEs with fainter Ly$\alpha$ luminosities (log $(L_{\rm Ly\alpha}/{\rm [erg\ s^{-1}]})<43$) than those of LABs. These models may not well explain the physical origin of the luminous and extended Ly$\alpha$ emission of LABs. 

In this paper, we present the identification of 2 new LABs at $z=4.9$ and 7.0. Including 5 LABs at $z=5.7$ and 6.6 identified by previous studies, we investigate the photometric and spectroscopic properties of a total of 7 LABs. We perform profile fitting to model the diffuse Ly$\alpha$ emission around LABs, and compare our best-fit models of LABs with those of LAHs in the literature. We investigate AGN activities in LABs with X-ray data and UV-line ratio diagnostics, and discuss the possible physical orgins of the extended Ly$\alpha$ emission around LABs.

This paper is organized as follows. In Section 2, we describe the observations, data, and identification of 2 new LABs. In Section 3, we present the spectroscopic analysis of LABs. Our results are shown in Section 4, and the discussions are presented in Section 5. We summarize our findings in Section 6. Throughout this paper, we use AB magnitudes \citep{oke1983} and physical distances unless we indicate otherwise. A $\Lambda$CGM cosmology with $\Omega_m=0.3$, $\Omega_\Lambda=0.7$, and $h=70$ is adopted.

\section{Observations, Data, and LAB Identification} \label{sec:obs}

In this paper, we study a total of 7 LABs that include 2 new LABs from our observations and 5 LABs from previous studies. We use photometric and spectroscopic data either from our observations or in the literature.

\subsection{New LABs Identified by CHORUS Survey}

\subsubsection{Imaging Observations and Data Reduction}

We carried out narrowband imaging observations with Subaru/Hyper Suprime-Cam (HSC; \citealt{miyazaki2018, komiyama2018, kawanomoto2018, furusawa2018}) in the course of project Cosmic HydrOgen Reionization Unveiled with Subaru (CHORUS; PI: A. K. Inoue). We used two narrowbands of NB718 ($\lambda_{\mathrm{c}}=7170$ \AA, FWHM=110 \AA) and NB973 ($\lambda_{\mathrm{c}}=9715$ \AA, FWHM=100 \AA). The central wavelengths of NB718 and NB973 filters were chosen to detect redshifted Ly$\alpha$ emission at $z=4.9$ and $7.0$, respectively. The NB718 data were taken on February 25, March 23, and March 25, 2017, while the NB973 observations were conducted on January 26 and 28, 2017. The NB718 observations covered the COSMOS field, and the NB973 observations were carried out in the SXDS and COSMOS fields. The effective survey areas were 1.64 and 1.50 deg$^{2}$ in the COSMOS and SXDS fields, respectively. The typical seeing sizes during observations were $0\farcs6-0\farcs9$.

We reduce the NB718 and NB973 images with the HSC pipeline (\citealt{bosch2018}) that uses codes from the Large Synoptic Survey Telescope (LSST) pipeline \citep{ivezic2008, axelrod2010, juric2015}. The astrometry and photometry are calibrated with the imaging data from the Panoramic Survey Telescope and Rapid Response System 1 (Pan-STARRS1; \citealt{schlafly2012, tonry2012, magnier2013}) survey. We do not use exposures with seeing sizes larger than $0\farcs9$ during the reduction because these exposures were taken under bad weather conditions. The total integration times of the reduced NB718 and NB973 images are 6.3 hours and 14.7 hours, respectively. The typical $5\sigma$ limiting magnitudes in a $1\farcs5$-diameter aperture are 26.2 in NB718 data and 24.9 in NB973 data.

During the reduction, in addition to the NB718 and NB973 images we also use the ultra-deep layer data of broadbands ($g$, $r$, $i$, $z$, and $y$) from the Subaru Strategic Program (SSP) survey (PI: S. Miyazaki; \citealt{aihara2018}) for source detection and forced photometry. Details of the source detection and forced photometry are described in \citet{bosch2018}. We do not use areas contaminated either by halos of bright stars (\citealt{coupon2018}) or low signal-to-noise ratio pixels such as field edges. The catalogs produced in this procedure are referred to as source catalogs in the following sections.

\subsubsection{Photometric Samples of LAEs at $z=4.9$ and 7.0}

To select LAEs at $z=4.9$, we use source catalogs of NB718, $g$, $r$, and $i$ filters. We apply the following color criteria,
\begin{equation} \label{eq:1}
\begin{gathered}
ri-NB718>0.7\ \mathrm{and}\ r-i>0.8\ \mathrm{and}\\
ri-NB718>(ri-NB718)_{3\sigma}\ \mathrm{and}\\
NB718^{\rm ap}<NB718_{5\sigma}^{\rm ap}\ \mathrm{and}\ g^{\rm ap}\geq g_{2\sigma}^{\rm ap},
\end{gathered}
\end{equation}
where the superscription "ap" indicates the aperture magnitude in a $2\farcs0$ diameter, and no superscription corresponds to the total magnitude. The total magnitude is measured by the CModel photometry described in \citet{bosch2018}. The $2\sigma$, $3\sigma$, and $5\sigma$ subscriptions stand for 2 sigma, 3 sigma, and 5 sigma detection limits, respectively. In Equation \ref{eq:1}, $ri$ is calculated by the linear combination of the fluxies in $r$ band $f_{\rm r}$ and $i$ band $f_{\rm i}$, following $f_{\rm ri}=0.3f_{\rm r}+0.7f_{\rm i}$. The $3\sigma$ error of the $ri-NB718$ color is given by $(ri-NB718)_{3\sigma}=-2.5\log(1\pm3\sqrt{f_{\rm{err},ri}^2+f_{\rm{err},\rm{NB718}}^2}/f_{\rm NB718})$, where $f_{\rm{err},ri}$ and $f_{\rm{err},\rm{NB718}}$ are the $1\sigma$ errors in $ri$ and $NB718$, respectively. These criteria allow us to choose LAEs with rest-frame Ly$\alpha$ equivalent widths (EW$_{0}$) greater than 10$\text{\AA}$.

\begin{figure}[ht]
\plotone{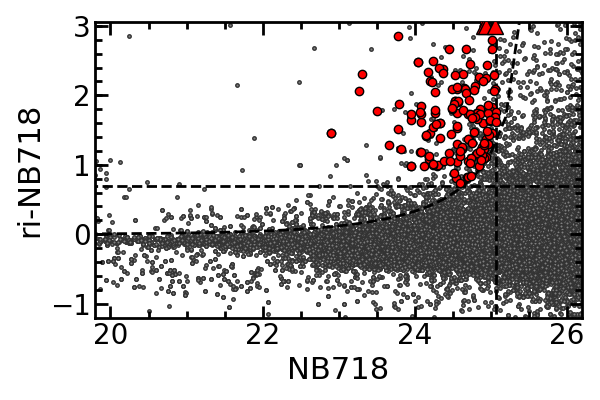}
\caption{Color-magnitude diagram of our LAE candidates at $z=4.9$. The red filled circles and triangles are the selected 141 LAE candidates. The triangles denote the LAEs whose $ri$-NB718 colors are greater than 3. The black filled circles are the other objects (including spurious sources) in our source catalog. The dashed lines present the color criteria in Equation \ref{eq:1}.} 
\label{fig:1}
\end{figure}

In total, 727 objects meet the color criteria. We then visually inspect these objects and exclude 586 spurious sources such as satellite trails. Finally we obtain 141 LAE candidates at $z=4.9$.
Figure \ref{fig:1} shows the color-magnitude diagram of our LAE candidates at $z=4.9$. 

The selection of LAEs at $z=7.0$ is presented in \citet{itoh2018}. Briefly, \citet{itoh2018} select LAEs with NB973 and broadbands ($g$,$r$,$i$,$z$, and $y$) following the criteria:
\begin{equation} \label{eq:2}
\begin{gathered}
\\
[(y^{\rm ap}<y_{3\sigma}^{\rm ap}\ \mathrm{and}\ y-NB973>1) \ \mathrm{or}\ y^{\rm ap}>y_{3\sigma}^{\rm ap}] \ \mathrm{and}\\
[(z^{\rm ap}<z_{3\sigma}^{\rm ap}\ \mathrm{and}\ z-y>2) \ \mathrm{or}\ z^{\rm ap}>z_{3\sigma}^{\rm ap}] \ \mathrm{and}\\ 
NB973^{\rm ap}<NB973_{5\sigma}^{\rm ap}\ \mathrm{and}\ g^{\rm ap}\geq g_{2\sigma}^{\rm ap}\ \mathrm{and}\\
r^{\rm ap}\geq r_{2\sigma}^{\rm ap}\ \mathrm{and}\ i^{\rm ap}\geq i_{2\sigma}^{\rm ap},
\end{gathered}
\end{equation}
where the meanings of superscriptions and subscriptions are the same as Equation \ref{eq:1}.

Finally, there are 34 LAE candidates at $z=7.0$ after we conduct the color selection and visual inspection.

\subsubsection{Identification of Two LABs}

\begin{figure}[ht]
\plotone{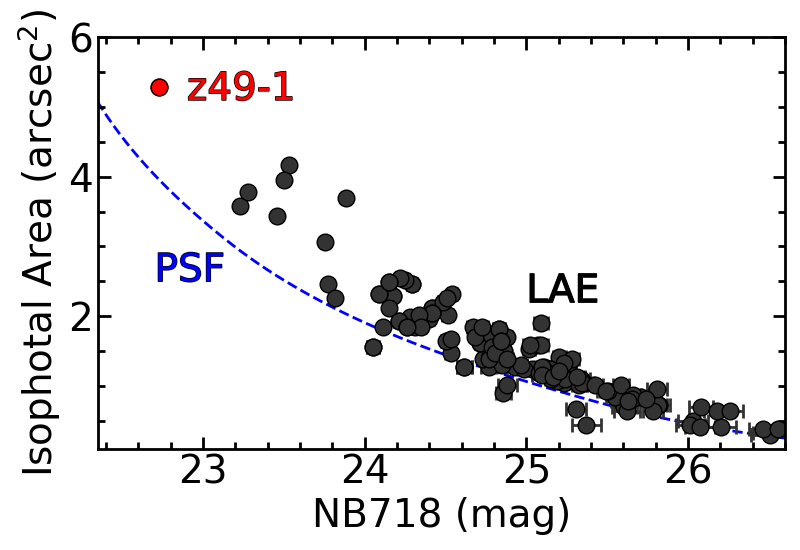}
\plotone{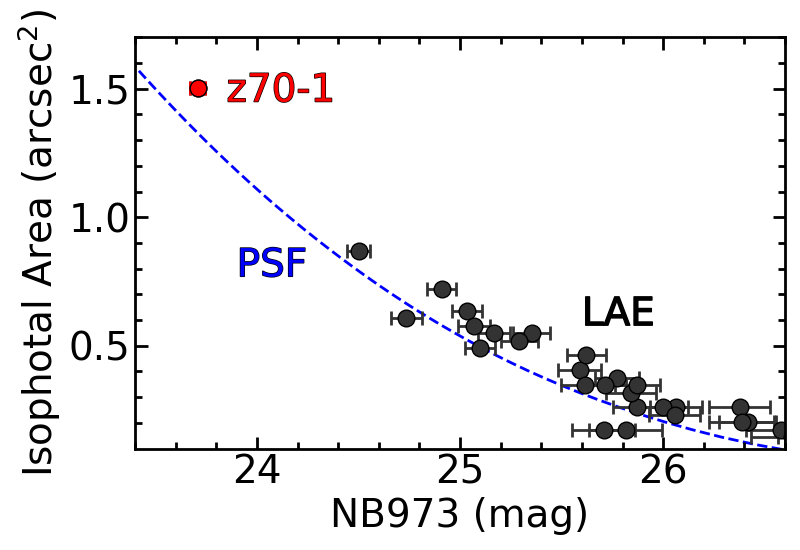}
\caption{Isophotal area as a function of NB718 (top) and NB973 (bottom) magnitudes for LAEs (black filled circles) at $z=4.9$ and $7.0$, respectively. The objects of z49-1 and z70-1 are indicated as red filled circles. The dashed lines show the size-magnitude relations of point sources.}
\label{fig:3}
\end{figure}

Figure \ref{fig:3} shows isophotal areas as a function of total narrowband magnitude for our LAE candidates at $z=4.9$ and 7.0. The isophotal area is defined as the area with a surface brightness above the 2$\sigma$ detection limit. We estimate the isophotal area-magnitude relations of point sources using the point spread functions (PSFs) in NB718 and NB973 images. Clearly, at $z=4.9$ and $7.0$ there are two very bright and large LABs that are named as
z49-1 (R.A.$=10^{\rm h}01^{\rm m}45.977$, decl.$=+2\degr02\arcmin44.28\arcsec$ [J2000]) and z70-1 (R.A.$=10^{\rm h}02^{\rm m}15.521$, decl.$=+2\degr40\arcmin33.23\arcsec$ [J2000]), respectively.  
The objects of z49-1 and z70-1 show large isophotal areas (157.5 and 42.2 physical kpc$^{2}$) and bright Ly$\alpha$ luminosities ($3.5\times10^{43}$ and $2.6\times10^{43}$ erg s$^{-1}$) that are distinguished from the other LAE candidates at each redshift. Snapshots of z49-1 and z70-1 are presented in Figure \ref{fig:4}.
The images of z49-1 and z70-1 from UltraVista ($Y$, $J$, $H$, and $K$ bands) and Spitzer/IRAC (3.6 $\mu$m and 4.5 $\mu$m bands) are shown in Figure \ref{fig:UVista_IRAC}, and will be analyzed in Section 4. 

\subsubsection{Spectroscopic Observations}
We carried out spectroscopic observations for z49-1 with Magellan/LDSS3 on May 28, 2017. The object of z49-1 was observed with an on-source exposure time of 1800s. The observations were conducted in the long-slit mode with a slit width of $2\farcs0$. We used the OG590 filter with the VPH-Red grism ($R\simeq680$) to cover the expected Ly$\alpha$ emission line at z=$4.9$. 

Spectroscopic observations for z70-1 were performed with Keck/DEIMOS \citep{faber2003} on Jan 6, 2019. The total on-source exposure time was 3.7 hours. However, we only used the data in the last 1.7 hours because the data in the first 2 hours were taken under bad weather conditions. The slit width was $1\farcs0$ during the observations in the multi-object spectroscopy (MOS) mode. The OG550 filter and the 830G grating ($R\simeq2900$ at 9700 \AA) were chosen to cover the wavelength where the Ly$\alpha$ emission line at z=$7.0$ was expected.

\subsection{LABs Identified in Previous Studies}

From previous studies we use 5 LABs including z57-1 and z57-2 (HSC J161927+551144 and HSC J161403+535701 in \citealt{shibuya2018a}) at z=5.7, z66-1 (Himiko in \citealt{ouchi2009}), z66-2 (CR7 in \citealt{sobral2015}), and z66-3 (HSC J100334+024546 in \citealt{shibuya2018a}) at $z=6.6$. The imaging data are available from the SSP survey and shown in Figure \ref{fig:4}. The spectra of z66-1, z66-2, and z66-3 are taken by \citet{ouchi2009}, \citet{sobral2015}, and \citet{shibuya2018a}, respectively. 

We carried out spectroscopic follow-up observations for z57-1 and z57-2 with Subaru/FOCAS on July 17, 2018. We chose a slit width of $0\farcs8$ in the MOS mode. The O58 filter and VPH900 grism ($R\simeq1500$) were used to cover the expected Ly$\alpha$ emission line at $z=5.7$. Finally, we obtained data with an on-source exposure time of 1200s for each target.

\subsection{Summary of Our LAB Samples}

\begin{figure}[ht]
\plotone{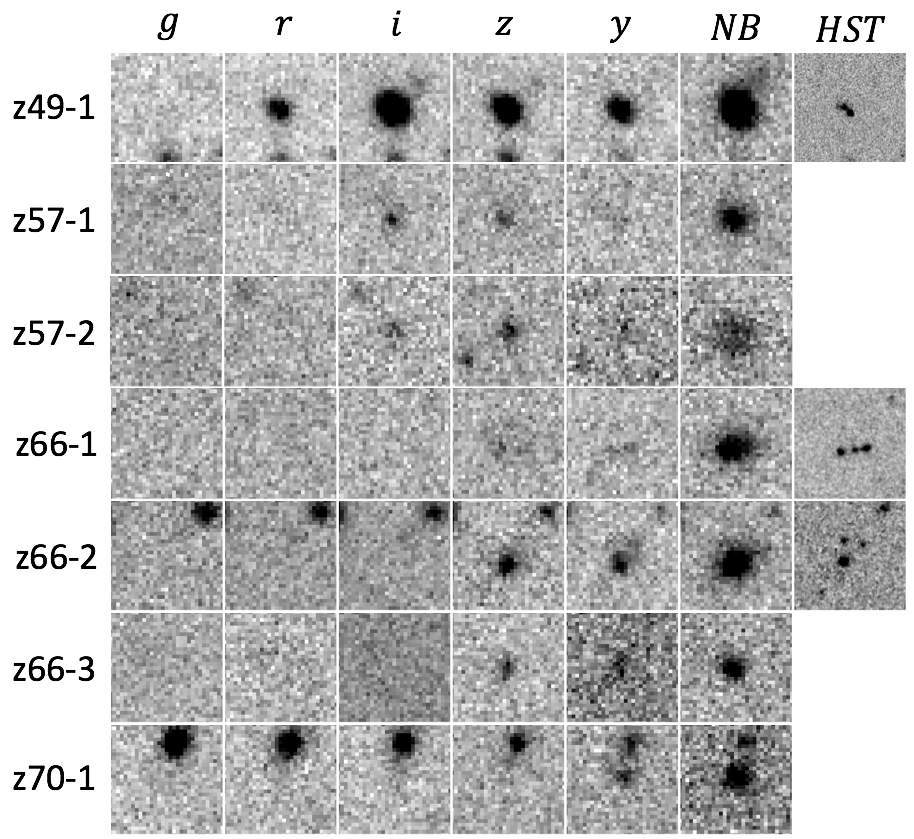}
\caption{Snapshots of the 7 LABs. The size of each image is $5\arcsec\times5\arcsec$. The HST corresponds to HST/WFC3 F814W, F125W, and F110W images for z49-1, z66-1 and z66-2, respectively.  }
\label{fig:4}
\end{figure}

Our final LAB samples include z49-1, z57-1, z57-2, z66-1, z66-2, z66-3 and z70-1 that are referred to as the 7 LABs in the following sections. From the snapshots of the 7 LABs in Figure \ref{fig:4}, we can see that apparently all of the 7 LABs are more extended in the narrowband images (NB718, NB816, NB921, and NB973) than the corresponding offband images ($i$, $z$, $y$, and $y$). Photometric properties of the 7 LABs are summarized in Table \ref{tab:1}. Details of spectroscopic observations of the 7 LABs are presented in Table \ref{tab:2}. The spectroscopic data will be shown and discussed in the next section.

\begin{figure}[ht]
\plotone{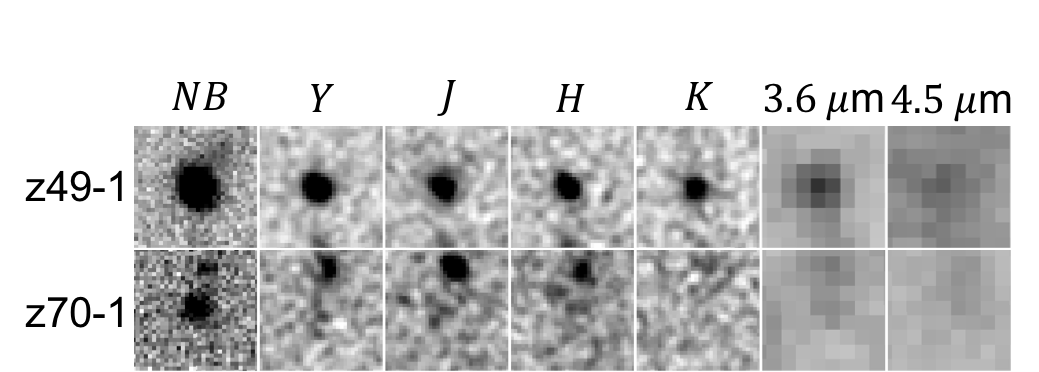}
\caption{Snapshots of z49-1 and z70-1 in narrowband, $Y$, $J$, $H$, $K$, 3.6 $\mu$m, and 4.5$\mu$m bands. The size of each image is $5\arcsec\times5\arcsec$. }
\label{fig:UVista_IRAC}
\end{figure}

\onecolumngrid

\begin{table}[ht]
\begin{center}
\caption{Photometric Properties of The 7 LABs}
\begin{tabular}{c  c  c  c  c  c  c  c}
\hline\hline
ID	& Object name	& Redshift	& $NB_{\mathrm{tot}}$ 	& $BB_{\mathrm{tot}}$	& log $L_{\mathrm{Ly\alpha}}$	& EW$_{0}$	&$\delta$ \\
 & & &(1) &(2) &(3) &($\text{\AA}$) &(4) \\
\hline
z49-1	&-		                    &4.9	&22.66	&23.89	&43.54	&47.5	&5.68 \\
z57-1	&HSC J161927+551144$^{\rm a}$	&5.7	&22.88	&24.86	&43.6	&71.4	&1.57 \\
z57-2	&HSC J161403+535701$^{\rm a}$	&5.7	&23.53	&25.32	&43.2	&20.6	&4.14 \\
z66-1	&Himiko$^{\rm b}$		        &6.6	&23.55	&25.00	&43.40	&78 	&2.09 \\
z66-2	&CR7$^{\rm c}$			        &6.6	&23.24	&24.92	&43.93	&211	&0.62 \\
z66-3	&HSC J100334+024546$^{\rm a}$ 	&6.6	&23.61	&24.97	&43.50	&61.1	&4.28 \\
z70-1	&-			                &7.0	&23.40	&25.09	&43.41	&73		&3.65 \\
\hline
\label{tab:1}
\end{tabular}
\end{center}
\hangindent=3cm 
\hspace{3cm}\textbf{Notes.}\\
Column 1: total narrowband magnitude in unit of mag. \\
Column 2: total broadband magnitude in unit of mag. \\
Column 3: photometric Ly$\alpha$ luminosity in unit of erg/s.\\
Column 4: LAE overdensity described in Section 4.2.\\
$^{\rm a}$ \citet{shibuya2018b}\\
$^{\rm b}$ \citet{ouchi2009}\\
$^{\rm c}$ \citet{sobral2015}\\
\end{table}

\begin{table}[ht]
\begin{center}
\caption{Summary of Spectroscopy}
\label{tab:2}
\begin{tabular}{ c  c  c  c  c  c  c }
\hline\hline
ID	&Instrument	&Filter	&Grism/grating	&Exp. time (s)	&Slit width ($\arcsec$)	&$z_{\text{spec}}$ \\
\hline
z49-1	        &Magellan/LDSS3	&OG590	&VPH-Red	&1,800	&2.0	&4.888 \\
z57-1			&Subaru/FOCAS	&O58	&VPH900		&1,200	&0.8	&5.709\\
z57-2			&Subaru/FOCAS	&O58	&VPH900		&1,200	&0.8	&5.733\\
z66-1$^{\rm a}$		&Keck/DEIMOS	&GG495	&830G		&10,800	&1.0	&6.595 \\
z66-2$^{\rm b}$		&VLT/X-SHOOTER	&-		&-			&8,100	&0.9	&6.604 \\
z66-3$^{\rm c}$ 	&Subaru/FOCAS	&O58	&VPH900		&6,000	&0.8	&6.575 \\
z70-1  	        &Keck/DEIMOS	&OG550	&830G		&6,000	&1.0	&6.965 \\
\hline
\end{tabular}
\end{center}
\hangindent=2.2cm 
\hspace{2.2cm}\textbf{Notes.}\\
$^{\rm a}$ \citet{ouchi2009}\\
$^{\rm b}$ \citet{sobral2015}\\
$^{\rm c}$ \citet{shibuya2018b}\\
\end{table}
\twocolumngrid

\section{Spectroscopic Analysis}

\begin{figure}[ht]
\epsscale{0.8}
\plotone{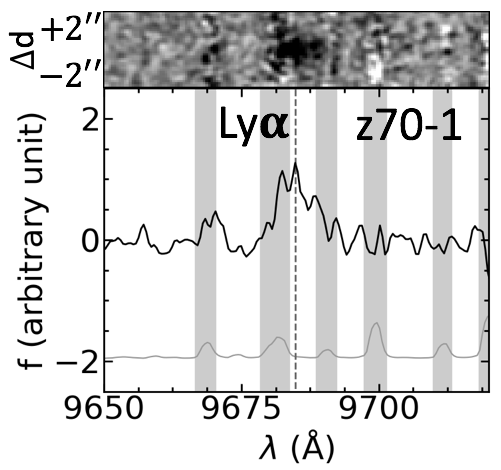}
\caption{Two-dimensional (top) and one-dimensional (bottom) spectra that show the Ly$\alpha$ emission (black solid line) of z70-1. The vertical dashed line indicates the Ly$\alpha$ line center. The gray solid line presents the sky emission lines. The gray shades represent the wavelength ranges with strong sky emission.} 
\label{fig:spec-LAB70}
\end{figure}

The spectrum of z70-1 is shown in Figure \ref{fig:spec-LAB70}. Because the emission line at 9686 $\mathrm{\AA}$ is partly overlapped by nearby sky lines, the line shape may be affected by the sky residual after sky subtraction. This emission line cannot be explained by an {\sc Oii} doublet, because the two peaks of an {\sc Oii} doublet at this wavelength would have a separation of $\sim 8$ $\mathrm{\AA}$ that is broader than the line observed. We find no other emission lines between $\sim$6000 and 10000 $\mathrm{\AA}$ that indicate a foreground source. We conclude that z70-1 is not likely a low-$z$ object but a LAB at $z=6.965$.

\begin{figure}[ht]
\epsscale{1.17}
\plotone{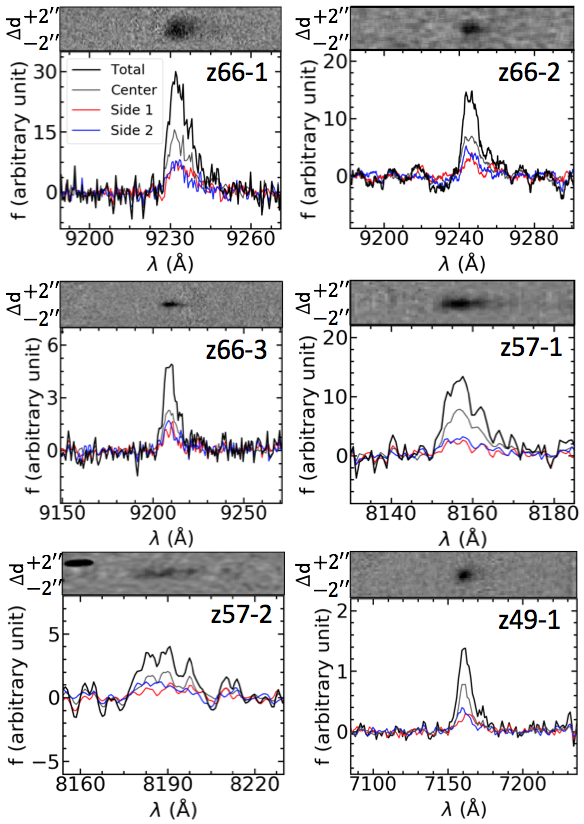}
\caption{Spectra of z49-1, z57-1, z57-2, z66-2, z66-1, and z66-3 that show the Ly$\alpha$ emission lines (black solid lines). In each panel, the two-dimensional spectrum is shown in the top and the one-dimensional spectrum is presented in the bottom. The center (gray), side 1 (red), and side 2 (blue) components are measured at positions with $\Delta d<0$, $\Delta d=0$, and $\Delta d>0$, respectively. The widths of the extraction slits are chosen arbitrarily to let the center, side 1, and side 2 components contain 50$\pm5$\%, 25$\pm5$\%, and 25$\pm5$\% of the total flux, respectively.} 
\label{fig:spec-lya}
\end{figure}

\begin{figure}[ht]
\epsscale{0.8}
\plotone{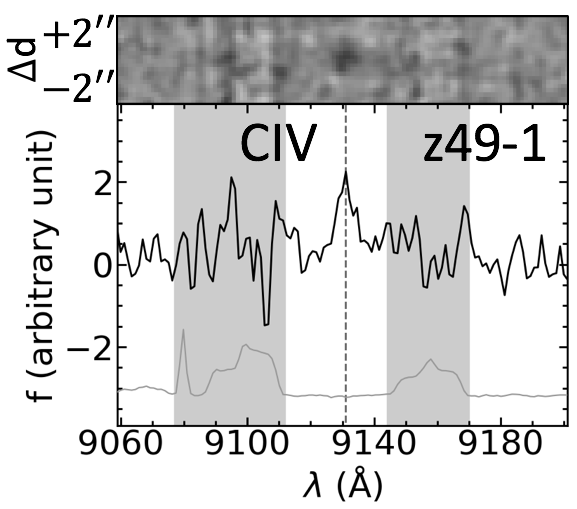}
\caption{Same as Figure \ref{fig:spec-LAB70}, but for the {\sc Civ} emission of z49-1.} 
\label{fig:spec-civ}
\end{figure}

Figure \ref{fig:spec-lya} presents the spectra of z49-1, z57-1, z57-2, z66-1, z66-2, and z66-3. The spectrum of z49-1 shows an emission line whose line center is at 7160 $\text{\AA}$. The line center is measured by fitting a gaussian function to the emission line. Additionally, on the spectrum we find another emission line whose line-center is at 9131 $\text{\AA}$, as presented in Figure \ref{fig:spec-civ}. These two emission lines can only be explained by an object emitting Ly$\alpha$ and {\sc Civ} lines simultaneously at $z=4.888$. The emission line at 7160 $\text{\AA}$ is asymmetric and has a red wing that is consistent with a high-$z$ Ly$\alpha$ emission line. The object of z49-1 is confirmed as a LAB at $z=4.888$. The Ly$\alpha$ and {\sc Civ} fluxes of z49-1 measured from the spectrum are $1.52\pm0.048 \times10^{-16}$ and $1.61\pm0.29 \times10^{-17}$ erg s$^{-1}$ cm$^{-2}$, respectively.

\begin{figure}[ht]
\plotone{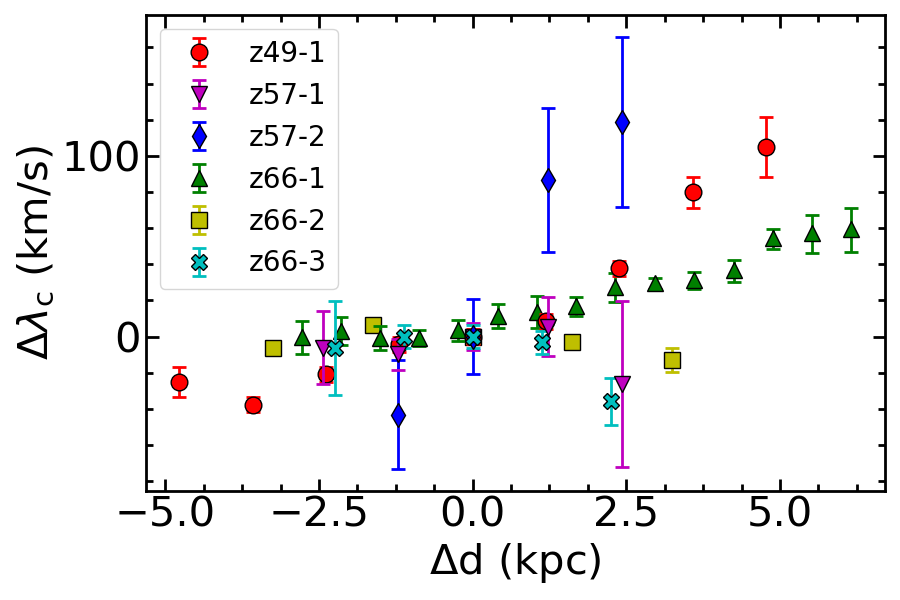}
\plotone{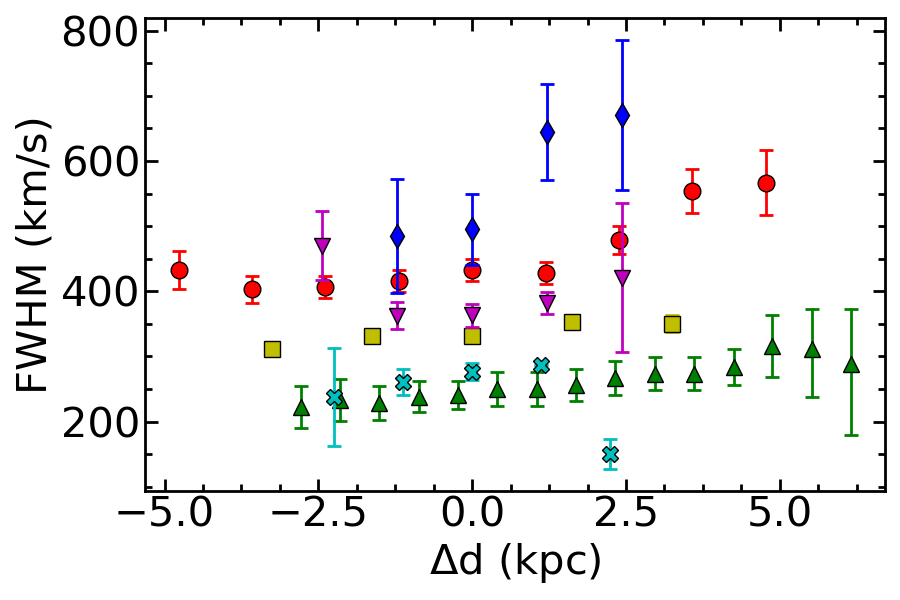}
\caption{Line-center offset $\Delta \lambda_{\mathrm{c}}$ (top) and FWHM (bottom) of the Ly$\alpha$ emission line as a function of positional offset $\Delta d$. The data of z66-1 are from Ouchi et al. (2009).} 
\label{fig:lc_fwhm}
\end{figure}

Figure \ref{fig:lc_fwhm} shows the line-center offset $\Delta \lambda_{\mathrm{c}}$ and FWHM of the Ly$\alpha$ emission line as a function of positional offset $\Delta d$. The $\Delta d$ is the distance between the position of a measurement and a Ly$\alpha$ source center. By definition, the Ly$\alpha$-source center is located at $\Delta d=0$. The positive direction of $\Delta d$ is from the blueshifted side to the redshifted side. The $\Delta \lambda_{\mathrm{c}}$ is calculated following $\Delta \lambda_{\mathrm{c}}=\lambda_{\mathrm{c}}(\Delta d)-\lambda_{\mathrm{c}}(0)$. Because the Ly$\alpha$ emission line of z70-1 is affected by nearby sky lines as we discussed earlier, we do not include z70-1 in this analysis. In Figure \ref{fig:lc_fwhm}, the $\Delta \lambda_{\mathrm{c}}$ has a positive correlation with $\Delta d$ although the correlation for z66-2 is weak. The correlation between $\Delta \lambda_{\mathrm{c}}$ and $\Delta d$ indicates velocity gradients in the the Ly$\alpha$ emission lines of our LABs. We notice that the FWHM also positively correlates with $\Delta d$. Clearly z49-1 and z57-2 have larger velocity gradients and FWHMs than the other LABs.

\section{Results}

\subsection{Ly$\alpha$ Surface Brightness Profiles}

To make Ly$\alpha$ images of the 7 LABs, we first match the PSFs of narrowband and offband images, and then subtract the offband images from corresponding narrowband images. We use a PSF matching method similar to the one discussed in \citet{aniano2011}. The PSF matching procedure is briefly described below. 

\begin{figure}[ht]
\epsscale{1.1}
\plotone{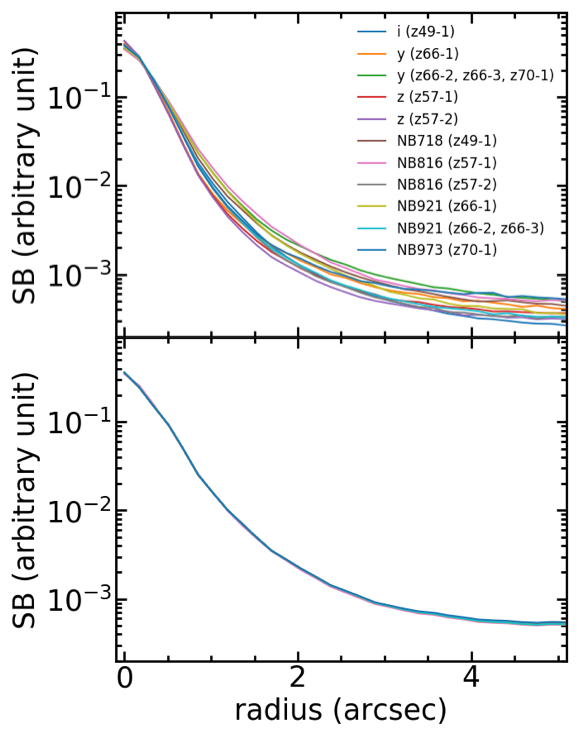}
\caption{PSFs before (top) and after (bottom) matching. The solid lines with different colors represent different PSFs. Each PSF is extracted from a $500\times500$ arcsec$^2$ region around each LAB in each filter. The names of filters and LABs are indicated in the legend. Because z66-2, z66-3, and z70-1 reside in the same field (COSMOS), we use the same $y$-band PSF for z66-2, z66-3, and z70-1. The same NB921 PSF is used for z66-2 and z66-3.} 
\label{fig:psf_match}
\end{figure}

First we extract the PSFs of narrowband and offband images by stacking 200-300 bright and unsaturated ($19<m_{\mathrm{AB}}<22$) point sources in each filter. These PSFs are referred to as initial PSFs. We choose the PSF with the largest FWHM among intial PSFs as the target PSF. Then we calculate convolution kernels that are used to convolve the initial PSFs to the target PSF by
\begin{equation}
K=\mathrm{FT}^{-1}\bigg(\mathrm{FT}(\mathrm{PSF}_{\rm t})\times\frac{1}{\mathrm{FT}(\mathrm{PSF}_{\rm i})}\bigg),
\end{equation}
where K, FT, FT$^{-1}$, PSF$_{\rm i}$, and PSF$_{\rm t}$ stand for the convolution kernel, Fourier transform, inverse Fourier transform, initial PSF, and target PSF, respectively. Finally we convolve the narrowband and offband images of the 7 LABs with the corresponding kernels to obtain PSF-matched images. The PSFs before and after matching are shown in Figure \ref{fig:psf_match}.

\begin{figure}[ht]
\epsscale{1.17}
\plotone{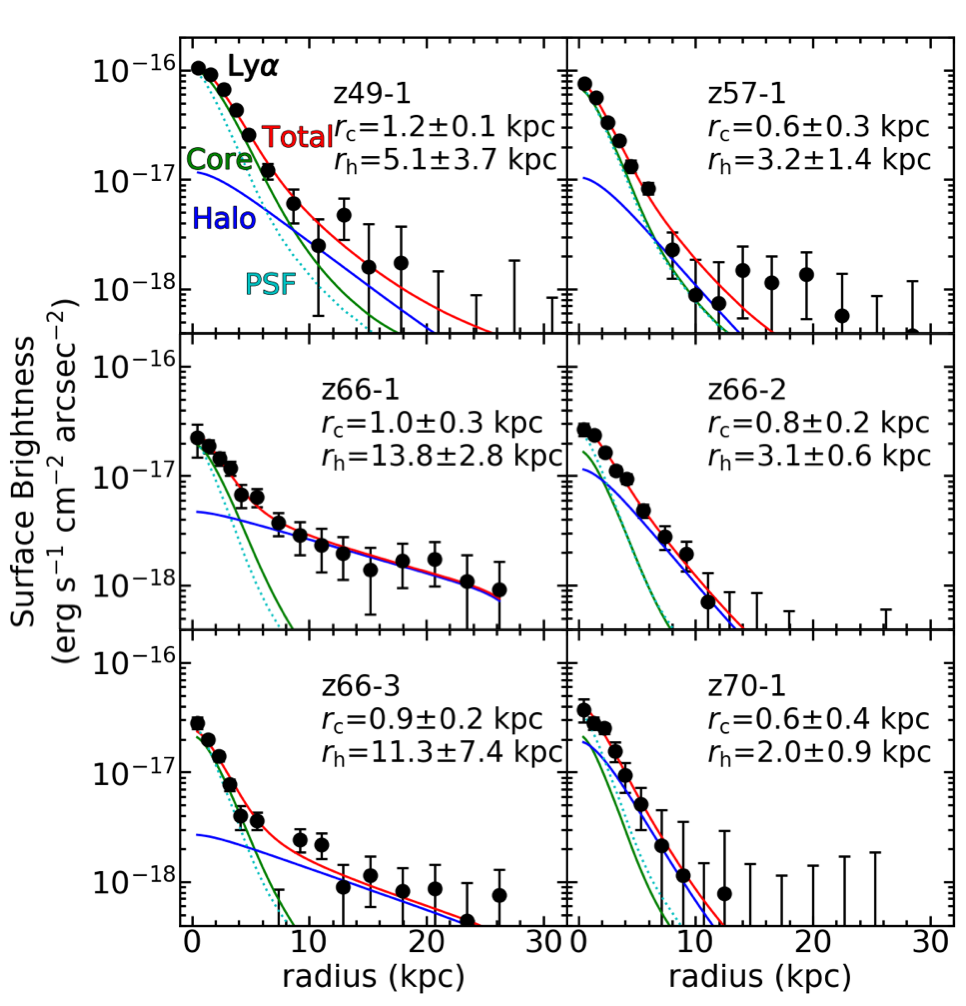}
\caption{Ly$\alpha$ surface brightness profiles of the 7 LABs except z57-2. The filled circles are the Ly$\alpha$ profiles extracted from the Ly$\alpha$ images. The red, green, and blue solid curves show the total, core, and halo best-fit models, respectively. The PSF is presented as a cyan dotted line.} 
\label{fig:fit}
\end{figure}

Figure \ref{fig:fit} shows Ly$\alpha$ surface brightness profiles $S_{\mathrm{Ly\alpha}}$ of the 7 LABs. To measure the scale lengths of the 7 LABs, we perform a two-component (core and halo) fitting that is similar to the one adopted by Leclercq et al. (2017). Specifically, we decompose the surface brightness profiles into core and halo components, following: 
\begin{equation} \label{eq:3}
\begin{gathered}
S_{\mathrm{cont}}(r)=\text{PSF}*A_{1}\exp(-r/r_{\mathrm{c}})\text{ and}\\
S_{\mathrm{Ly\alpha}}(r)=\text{PSF}*[A_{2}\exp(-r/r_{\mathrm{c}})+A_{3}\exp(-r/r_{\mathrm{h}})],\\
\end{gathered}
\end{equation}
where $r_{\mathrm{c}}$ and $r_{\mathrm{h}}$ are the scale lengths of core and halo components, respectively. The ``$*$" sign stands for convolution. The $A_{1}$, $A_{2}$, and $A_{3}$ are free parameters. The continuum profile $S_{\mathrm{cont}}$ is extracted from the offband images, while the Ly$\alpha$ profile $S_{\mathrm{Ly\alpha}}$ is measured in the Ly$\alpha$ images. We first fit $S_{\mathrm{cont}}$ with two free parameters $A_{1}$ and $r_{\mathrm{c}}$ to measure $r_{\mathrm{c}}$. Then we use this $r_{\mathrm{c}}$ value to fit $S_{\mathrm{Ly\alpha}}$ with three free parameters $A_{2}$, $A_{3}$, and $r_{\mathrm{h}}$ to measure $r_{\mathrm{h}}$. The errors of $S_{\mathrm{cont}}$ and $S_{\mathrm{Ly\alpha}}$ are considered in the fitting.

\begin{figure}[ht]
    \centering
    \plotone{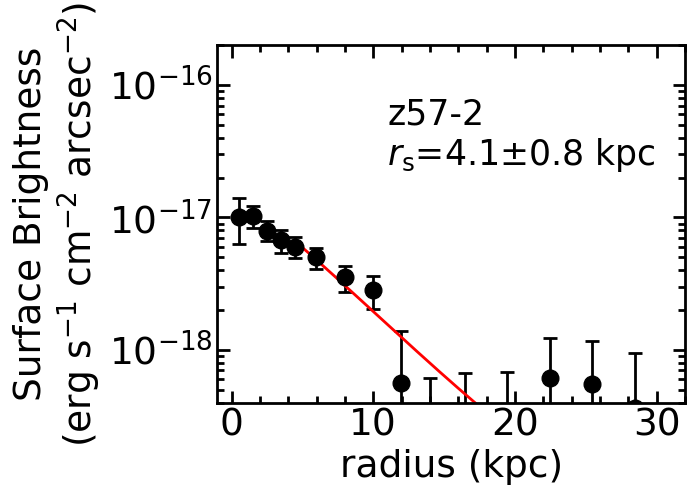}
    \caption{Same as Figure 6, but for z57-2. The red solid line represents the best-fit one-component exponential function.}
    \label{fig:Z57-2_fit}
\end{figure}

Figure \ref{fig:fit} shows the best-fit Ly$\alpha$ surface brightness profiles of our LABs. Because there is an offset between the positions of the Ly$\alpha$ and continuum centers of z57-2, we cannot perform the two-component fitting that requires the Ly$\alpha$ and continuum centers to be the same. Instead we use a one-component exponential function to fit the Ly$\alpha$ profile of z57-2 in the halo region ($r\gtrsim 5$ kpc), following
\begin{equation} \label{eq:4}
\begin{gathered}
S_{\mathrm{Ly\alpha}}(r)=\text{PSF}*[A\exp(-r/r_{\mathrm{s}})], \\
\end{gathered}
\end{equation}
where the meanings of the $S_{\mathrm{Ly\alpha}}$, PSF, and ``$*$" sign are the same as Equation \ref{eq:3}. The $A$ is a free parameter. The fitting result of z57-2 is shown in Figure \ref{fig:Z57-2_fit}.

\begin{figure}[ht]
\plotone{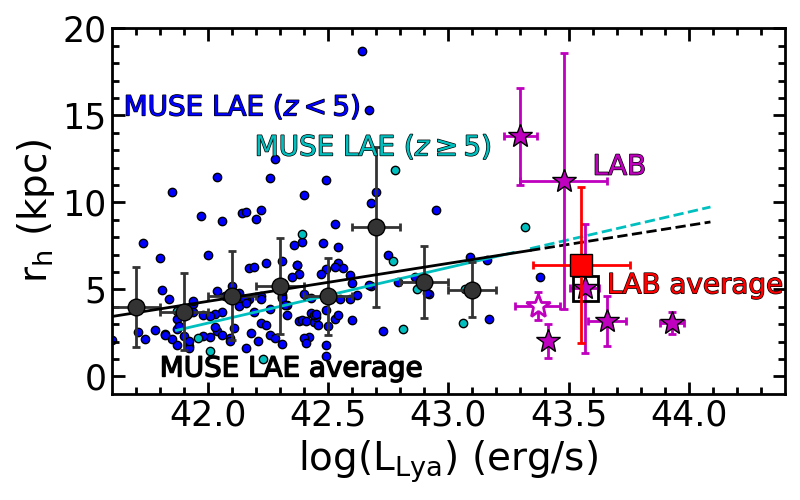}
\plotone{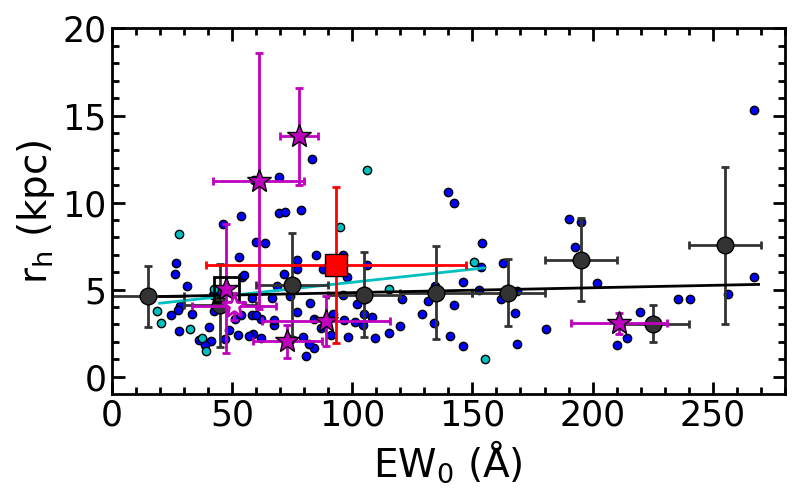}
\caption{Halo scale length as a function of Ly$\alpha$ luminosity (top) and Ly$\alpha$ rest-frame equivalent width (bottom) of the 7 LABs (star marks) and LAEs from Leclercq et al. (2017) (filled circles). The empty star represents z57-2 that does not have a two-component fitting result. The red filled square shows the average value of our LABs. The MUSE LAEs at $z<5$ and $z\geq5$ are blue and cyan filled circles, respectively. The average values of MUSE LAEs are shown as black filled circles. The solid lines represent the best-fit linear functions to the MUSE LAEs at $z=3-6$ (black) and $z=5-6$ (cyan), while the dashed lines are the extrapolations of the best-fit functions. It is clear that the average values of our LABs are consistent with the extrapolations of the best-fit functions of MUSE LAEs at $z=3-6$ and $z=5-6$. In the top panel, we slightly shift z49-1 (boxed star) along the horizontal axis by +0.03 to avoid overlaps.} 
\label{fig:r}
\end{figure}

\begin{figure}[ht]
    \centering
    \plotone{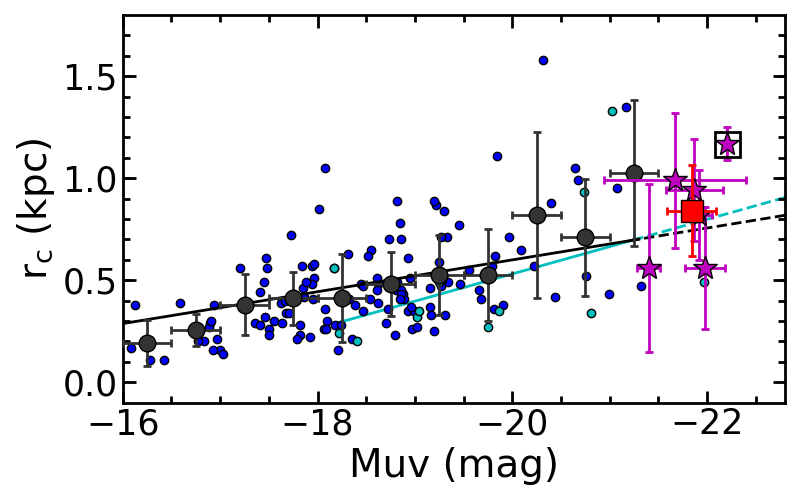}
    \plotone{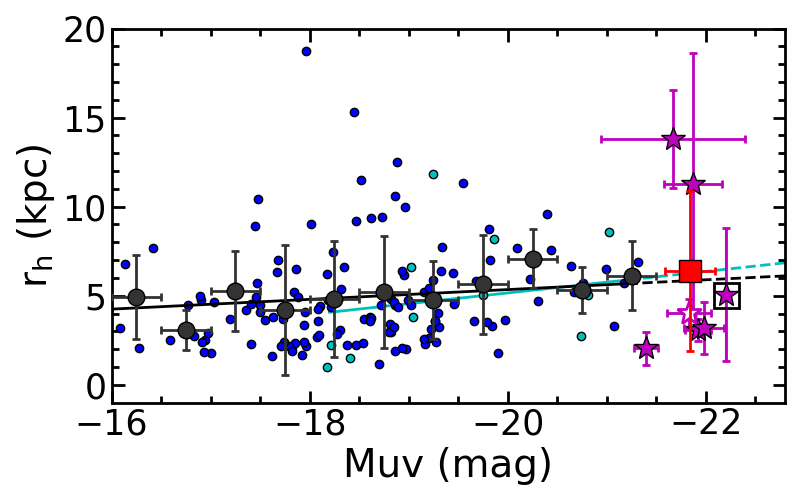}
    \caption{Same as Figure \ref{fig:r}, but for the core scale length (top) and halo scale length (bottom) as a function of continuum magnitude.}
    \label{fig:r2}
\end{figure}

\begin{figure}[ht]
    \centering
    \plotone{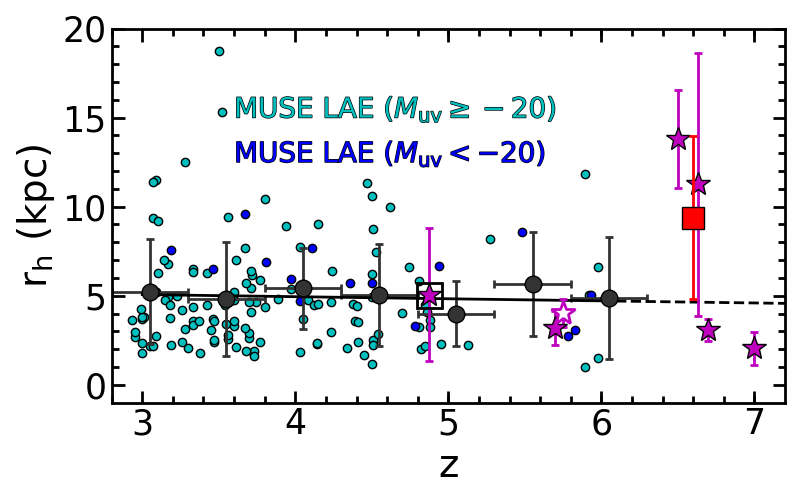}
    \caption{Halo scale length as a function of redshift. The MUSE LAEs with $M_{\rm UV}\geq-20$ and $M_{\rm UV}<-20$ are presented as cyan and blue filled circles, respectively. The meanings of stars and blacked filled circles are the same as Figure \ref{fig:r}. We use z66-1, z66-2, and z66-3 to calculate the LAB average value at $z=6.6$ (red filled square). The objects of z57-2, z66-1, z66-2, and z66-3 are slightly shifted along the horizontal axis by +0.05, -0.1, +0.1, and +0.03 to avoid overlaps, respectively.}
    \label{fig:r3}
\end{figure}

We compare the best-fit $r_{\mathrm{h}}$ values as a function of Ly$\alpha$ luminosities $L_{\mathrm{Ly\alpha}}$, Ly$\alpha$ rest-frame equivalent widths EW$_{0}$, continuum magnitudes $M_{\mathrm{UV}}$, and redshifts $z$ of the 7 LABs with those of LAEs from Leclercq et al. (2017), as shown in Figures \ref{fig:r}, \ref{fig:r2}, and \ref{fig:r3}.
When calculating the LAB average value, we do not use the best-fit $r_{\mathrm{h}}$ of z57-2 from the one-component exponential function fitting. In Figures \ref{fig:r} and \ref{fig:r2}, our LABs are consistent with the extrapolations of correlations between $r_{\mathrm{h}}$ and galaxy properties including $L_{\mathrm{Ly\alpha}}$, EW$_{0}$, and $M_{\mathrm{UV}}$ of the MUSE LAEs at $z=5-6$. This suggests that our LABs and MUSE LAEs have similar connections between the diffuse Ly$\alpha$ emission and properties of the host galaxies, and that typical LABs at $z\gtrsim5$ are not special objects, but star-forming galaxies at the bright end. We also find that our LABs are consistent with the positive correlation between the $r_{\mathrm{c}}$ as a function of $M_{\mathrm{UV}}$ of MUSE LAEs, which is expected from the size evolution discussed in \citet{shibuya2015} and \citet{shibuya2019}.

\citet{leclercq2017} find no significant evolution of the $r_{\mathrm{h}}$ of MUSE LAEs at $z=3-6$. Consistently, we notice that in Figure \ref{fig:r3} the $r_{\mathrm{h}}$ of our LABs does not evolve significantly between $z$=4.9 and 7.0. Moreover, our LABs fall on the extrapolations of correlations between $r_{\mathrm{h}}$ and galaxy properties of the MUSE LAEs at $z=3-6$ in Figures \ref{fig:r}, \ref{fig:r2}, and \ref{fig:r3}.

\subsection{Large Scale Structure around LABs}
\begin{figure}[ht]
    \centering
    \epsscale{0.9}
    \plottwo{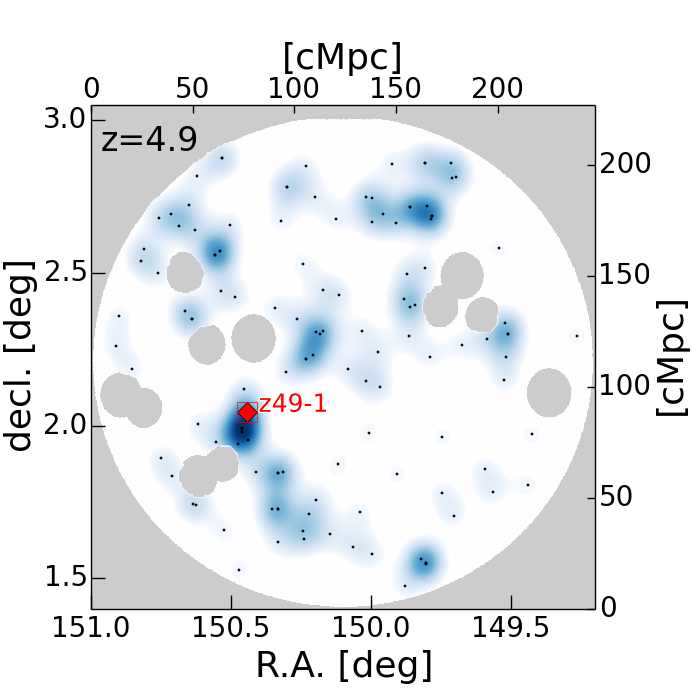}{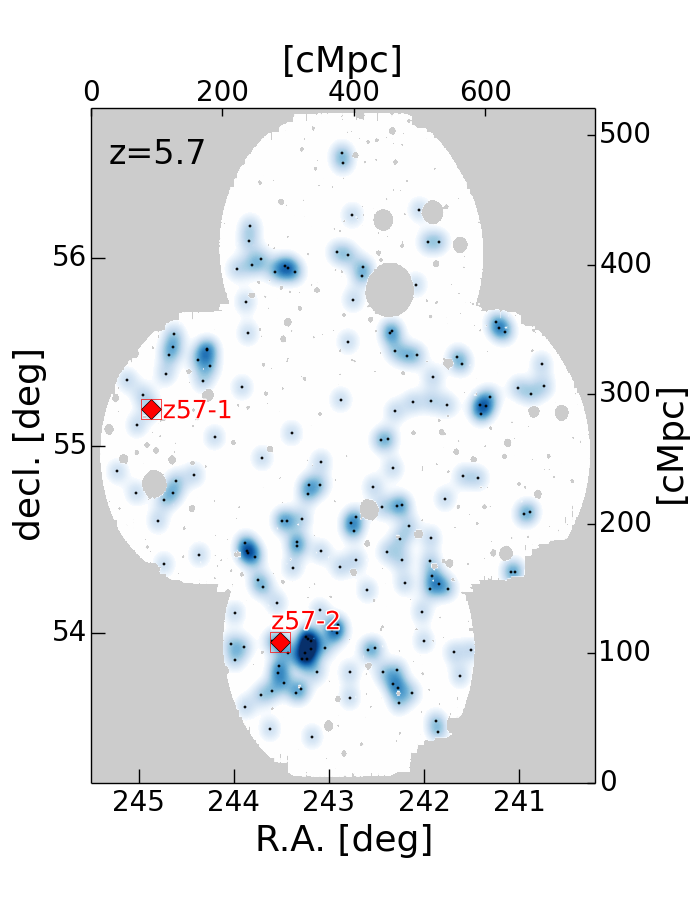}
    \plottwo{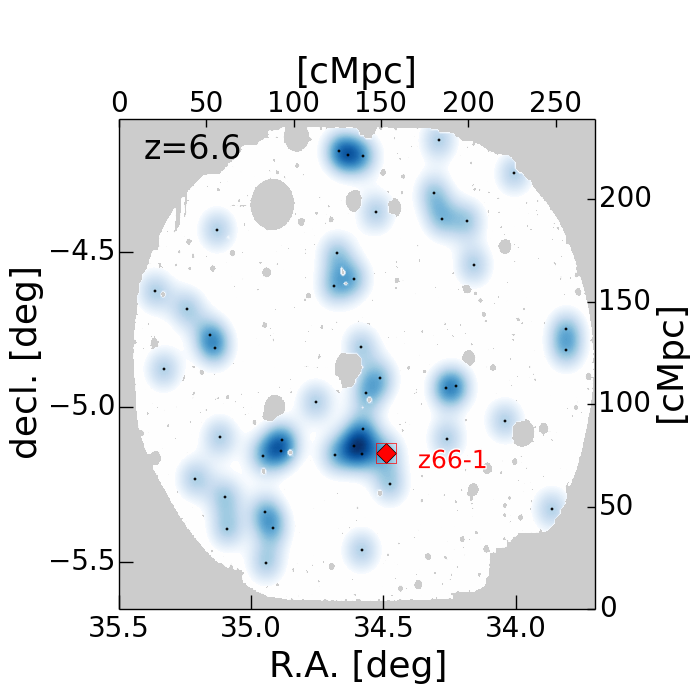}{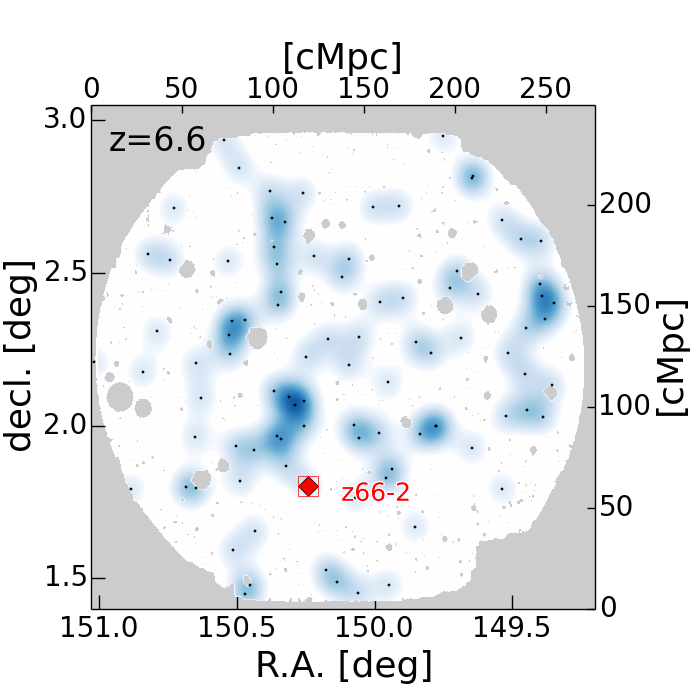}
    \plottwo{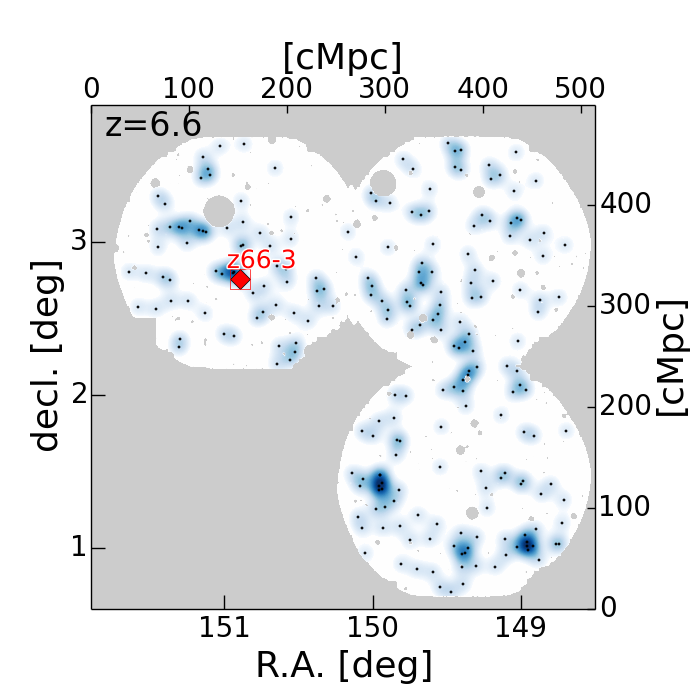}{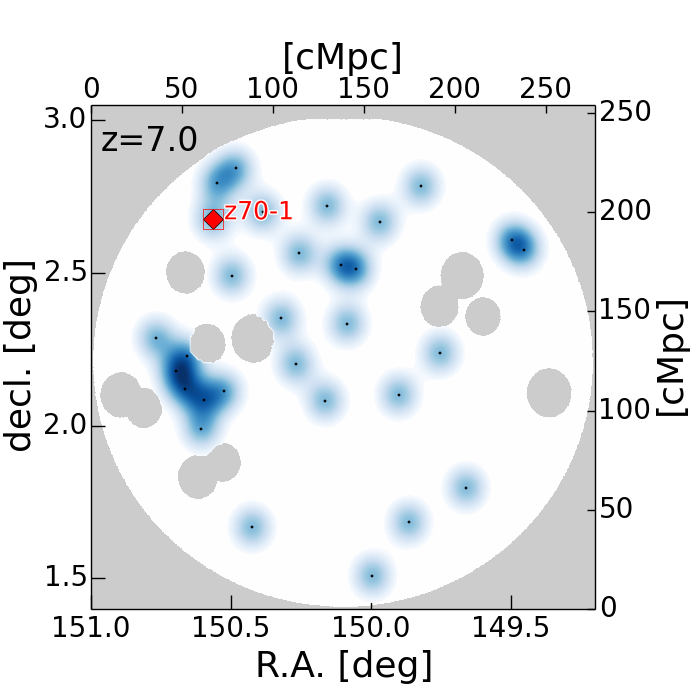}
    \caption{Overdensity maps of LAEs at $z=4.9$, 5.7, 6.6, and 7.0. The red diamonds indicate our LABs, while the other LAEs are shown as black dots. The blue contours present the number densities of LAEs. Dark blue regions have higher number densities than light blue regions.}
    \label{fig:overdensity}
\end{figure}

To investigate the large scale structure around our LABs, we calculate the LAE overdensity $\delta$ at $z=4.9$, 5.7, 6.6, and 7.0 in the same manner as in \citet{harikane2019}. The $\delta$ is defined as

\begin{equation} \label{eq:delta}
\begin{gathered}
\delta=\frac{n-\bar{n}}{\bar{n}}, \\
\end{gathered}
\end{equation}
where $n$ and $\bar{n}$ are the number and average number of LAEs in a cylinder, respectively. The radius of the cylinder is $\sim$10 comoving Mpc (cMpc). This radius is the typical size of protoclusters whose masses grow to $\sim10^{15}$ $M_{\odot}$ at $z=0$ in \citet{chiang2013}. The length of the cylinder is $\sim40$ cMpc, consistent with the redshift range of LAEs selected by narrowbands. Figure \ref{fig:overdensity} shows the overdensity maps of LAEs at $z=4.9$, 5.7, 6.6, and 7.0. The $\delta$ of each LAB is presented in Table \ref{tab:1}. \citet{kikuta2019} show that most of their LABs reside in overdensed regions at $z\sim3$. Consistently, we find that all the 7 LABs except z66-2 are located in overdensed regions. 

\onecolumngrid

\twocolumngrid

\begin{figure}[ht]
    \centering
    \plotone{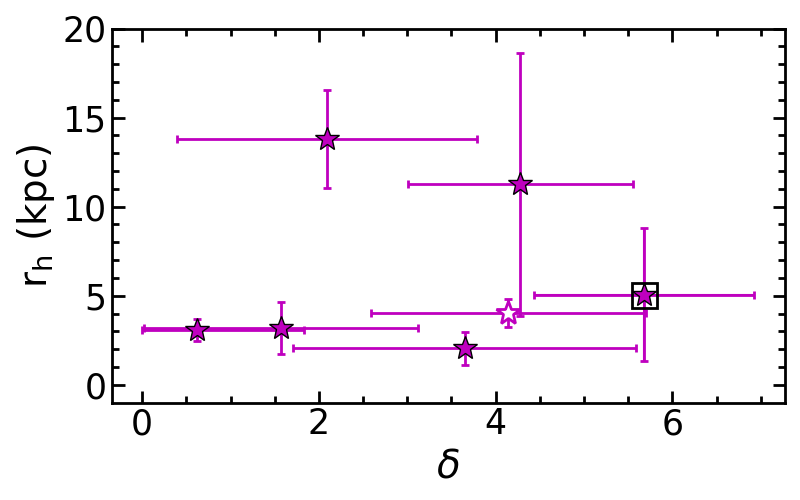}
    \caption{Halo scale length as a function of LAE overdensity of our LABs. The meanings of symbols are the same as those in Figure \ref{fig:r}.}
    \label{fig:rh-delta}
\end{figure}

 Figure \ref{fig:rh-delta} shows the $r_{\mathrm{h}}$ as a function of $\delta$ of our LABs. To test the correlation between the $r_{\mathrm{h}}$ and $\delta$, we calculate the Spearman's rank correlation coefficient $\rho$ to be 0.43 with a $p$-value of 0.34. We do not consider the errors of $r_{\mathrm{h}}$ and $\delta$ when calculating the $\rho$ and $p$-value. Although \citet{matsuda2012} find a positive correlation between the halo scale length and LAE overdensity of LAEs at $z=3.1$, our correlation test suggests that there is no significant correlation between the $r_{\mathrm{h}}$ and $\delta$ of our LABs at $z=4.9$-7.0.

\subsection{AGN Activity}
Because the bright Ly$\alpha$ luminosities ($>10^{43.4}$ erg s$^{-1}$) of the 7 LABs make them possible hosts of AGNs, we investigate the AGN activities in LABs with X-ray and spectroscopic data. None of the 7 LABs have X-ray counterparts in images and catalogs of XMM/Newton and Chandra in the literature \citep{scoville2007,hasinger2007,civano2016,marchesi2016}. The spectra of the 7 LABs do not show {\sc Nv} emission indicative of AGNs. 

\citet{shibuya2018b} investigate 21 bright LAEs that are not broad-line AGNs at $z=6-7$, and find that the LAEs have Ly$\alpha$ line widths of $\sim200-400$ km s$^{-1}$. Consistently, z57-1, z66-1, z66-2, z66-3, and z70-1 also show Ly$\alpha$ line widths of $\sim200-400$ km s$^{-1}$ in Figure \ref{fig:lc_fwhm}, suggesting that z57-1, z66-1, z66-2, z66-3, and z70-1 are not broad-line AGNs. On the other hand, the Ly$\alpha$ line widths of z49-1 and z57-2 are systematically larger than 400 km s$^{-1}$. Because z49-1 has a very clear continuum center that z57-2 does not show in Figure \ref{fig:4}, it is possible that a hidden AGN is the origin of the relatively large Ly$\alpha$ line width of z49-1. The large Ly$\alpha$ line width of z57-2 is not likely caused by an AGN, but by mergers or dense neutral hydrogen gas in the {\sc Hi} region.

\begin{figure}[ht]
    \centering
    \plotone{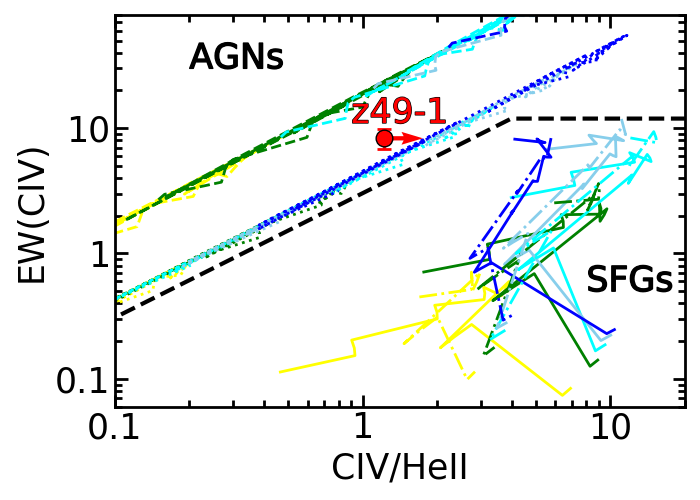}
    \caption{{\sc Civ} equivalent width as a function of {\sc Civ}/He{\sc ii} ratio. The object z49-1 is represented as a red filled circle. We show two AGN models with power-law indices of $\alpha=-2.0$ (dotted line) and -1.2 (dashed line), and two SFG models of POPSTAR (solid line) and BPASS (dash-dotted line) from \citet{nakajima2018}. The ionization parameters $\log U$ of the AGN and SFG models are -2.5 (yellow), -2.0 (green), -1.5 (cyan), -1.0 (skyblue), and -0.5 (blue). The black dashed line represents the threshold that distinguishes between AGNs and SFGs.}
    \label{fig:AGN_SFG}
\end{figure}

In Section 3, we show that z49-1 has a {\sc Civ} emission line with a line width of $317\pm132$ km/s. The rest-frame equivalent width of the {\sc Civ} emission is $8.3\pm1.5$\AA. The spectrum shows no He{\sc ii} emission above the 2$\sigma$ detection limit. We use the 2$\sigma$ detection limit as an upper limit of the He{\sc ii} flux, and find that the lower limit of the {\sc Civ} to He{\sc ii} ratio is $\sim1.2$. We compare the {\sc Civ} reset-frame equivalent width and {\sc Civ} to He{\sc ii} ratio with the AGN and star-forming galaxy (SFG) models in \cite{nakajima2018}, and find that the z49-1 is consistent with both the AGN and low-metallicity SFG models (Figure \ref{fig:AGN_SFG}). This result indicates that z49-1 is a candidate of a high-$z$ AGN, although the possibility of a low-metallicity SFG cannot be ruled out.

As we discussed in Section 1, AGNs have been identified in all of the LAEs with bright Ly$\alpha$ luminosities (log $(L_{\rm Ly\alpha}/{\rm [erg\ s^{-1}]})\gtrsim43.4$) at $z\sim2-3$ in \citet{konno2016} and \citet{sobral2018}. Similary, \citet{overzier2013} show that at least 63\% of LABs at $z\sim2-3$ are associated with luminous AGNs. On the other hand, no AGN has been confirmed to exist in LABs at $z\gtrsim5$ including our LABs. This may suggest that typical LABs at $z\gtrsim5$ are less likely to be powered by luminous AGNs than LABs at $z\sim2-3$.

\subsection{Stellar Population}\label{sec:SED}

\begin{figure}[ht]
\plotone{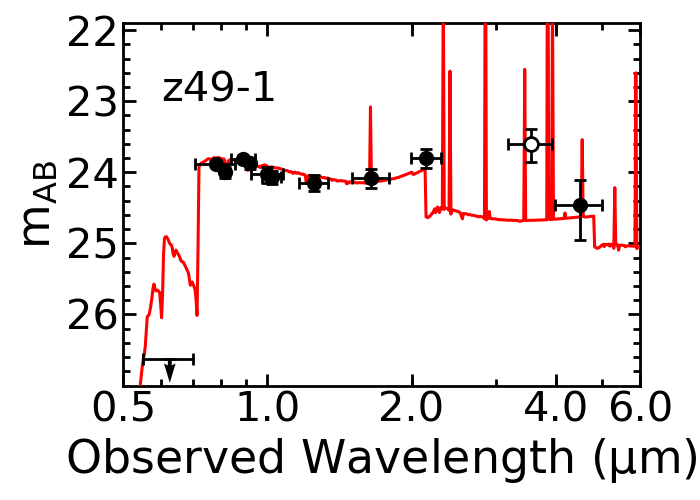}
\plotone{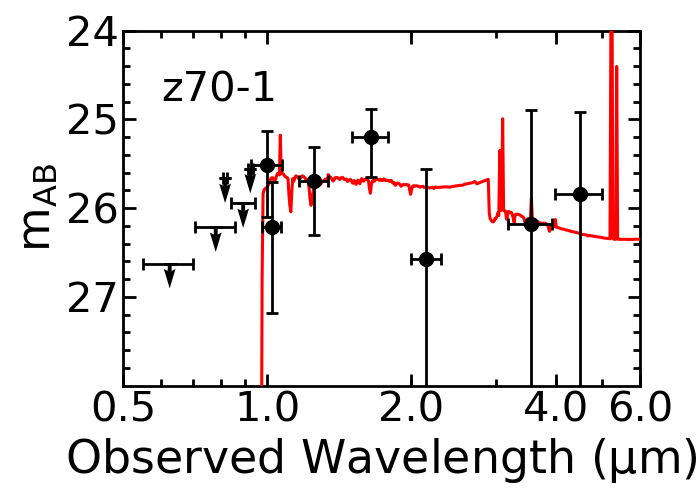}
\caption{SEDs of the best-fit models of z49-1 (top) and z70-1 (bottom). The SED models are presented as the red solid curves. The black filled circles are total magnitudes measured in $g$, $r$, $i$, $z$, $y$, NB816, NB921, $Y$, $J$, $H$, $K$, 3.6 $\mu$m, and 4.5 $\mu$m bands. The black open circle indicates the 3.6$\mu$m band photometry that we do not use in the SED fitting of z49-1. The horizontal error bars represent the filter bandwidths. The vertical error bars show the 1$\sigma$ errors in magnitude. The arrows indicate 3$\sigma$ upper limits.}
\label{fig:SED}
\end{figure}

We perform SED fitting on z49-1 and z70-1 using total magnitudes measured in Subaru HSC ($g$, $r$, $i$, $z$, $y$, NB816, and NB921), UltraVista ($Y$, $J$, $H$, and $K$), and Spitzer/IRAC (3.6 $\mu$m and 4.5 $\mu$m bands) images. In our SED fitting, we consider the contributions from both nebular and stellar populations. The nebular spectra (emission lines and continua) are calculated basically following \citet{schaerer2009}. We use the stellar population synthesis model GALAXEV \citep{bc2003} with Salpeter's initial mass function \citep{salpeter1955} to obtain stellar SEDs. A constant star formation history is assumed. Details of our SED fitting method are discribed in \citet{ono2010}. Because the 3.6$\mu$m band is contaminated by H$\alpha$ emission at $z=4.9$, we do not use the photometry of the 3.6$\mu$m band in our SED fitting of z49-1. The best-fit SEDs of z49-1 and z70-1 are shown in Figure \ref{fig:SED}. The properties of the best-fit SEDs are summarized in Table \ref{tab:3}.

\onecolumngrid

\begin{table}[ht]
\begin{center}
\caption{Properties of best-fit SEDs}
\label{tab:3}
\begin{tabular}{ c  c  c  c  c  c  }
\hline\hline
ID	            &$Z$	&$\log M_{*}$	&$E(B-V)_{*}$	&$\log({\rm Age})$	&$\log({\rm SFR})$ \\
                &($Z_{\odot}$)  &($M_{\odot}$)   &(mag)      &(yr)       &($M_{\odot}$ yr$^{-1}$) \\
\hline
z49-1	        &0.004	&$9.0^{+0.2}_{-0.1}$	 &0.05	&$6.6^{+0.5}_{-1.5}$	&$2.4^{+1.4}_{-0.3}$ \\
z66-1$^{\rm a}$ &0.2    &$10.18^{+0.05}_{-0.07}$ &0.15  &$8.26^{+0.05}_{-0.05}$ &$2.00^{+0.01}_{-0.01}$\\
z66-2$^{\rm b}$ &0.005-0.2 &$\sim10.3$         &0.0-0.5 &$\sim8.8$              &$\sim1.4$\\
z70-1  	        &0.02	&$<9.1$	                 &0.10	&$<7.7$	                &$2.0^{+1.8}_{-0.8}$ \\
\hline
\end{tabular}
\end{center}
\hangindent=4cm 
\hspace{4cm}\textbf{Notes.}\\
$^{\rm a}$ best-fit SED from \citet{ouchi2013}\\
$^{\rm b}$ best-fit SED from \citet{sobral2015}\\
\end{table}
\twocolumngrid

\subsection{H$\alpha$ Emission of z49-1}

The 3.6 $\mu$m image of z49-1 in Figure \ref{fig:UVista_IRAC} shows a clear color excess that is caused by the redshifted H$\alpha$. Comparing with the best-fit SED obtained in Section \ref{sec:SED}, we measure the observed 3.6 $\mu$m excess that corresponds to a H$\alpha$ luminosity of $(3.6\pm1.2)\times10^{43}$ erg s$^{-1}$. Assuming the case-B recombination and no dust extinction suggested by the best-fit SED, we estimate the expected Ly$\alpha$ luminosity to be $(3.2\pm1.1)\times10^{44}$ erg s$^{-1}$. The Ly$\alpha$ escape fraction is the observed Ly$\alpha$ luminosity divided by the expected Ly$\alpha$ luminosity, $(3.5\times10^{43})$/$(3.2\times10^{44})$=0.11$\pm$0.04.

\section{Discussion}
\subsection{Identification of the Most Distant LAB at $z=7.0$}
\begin{figure}[ht]
    \centering
    \plottwo{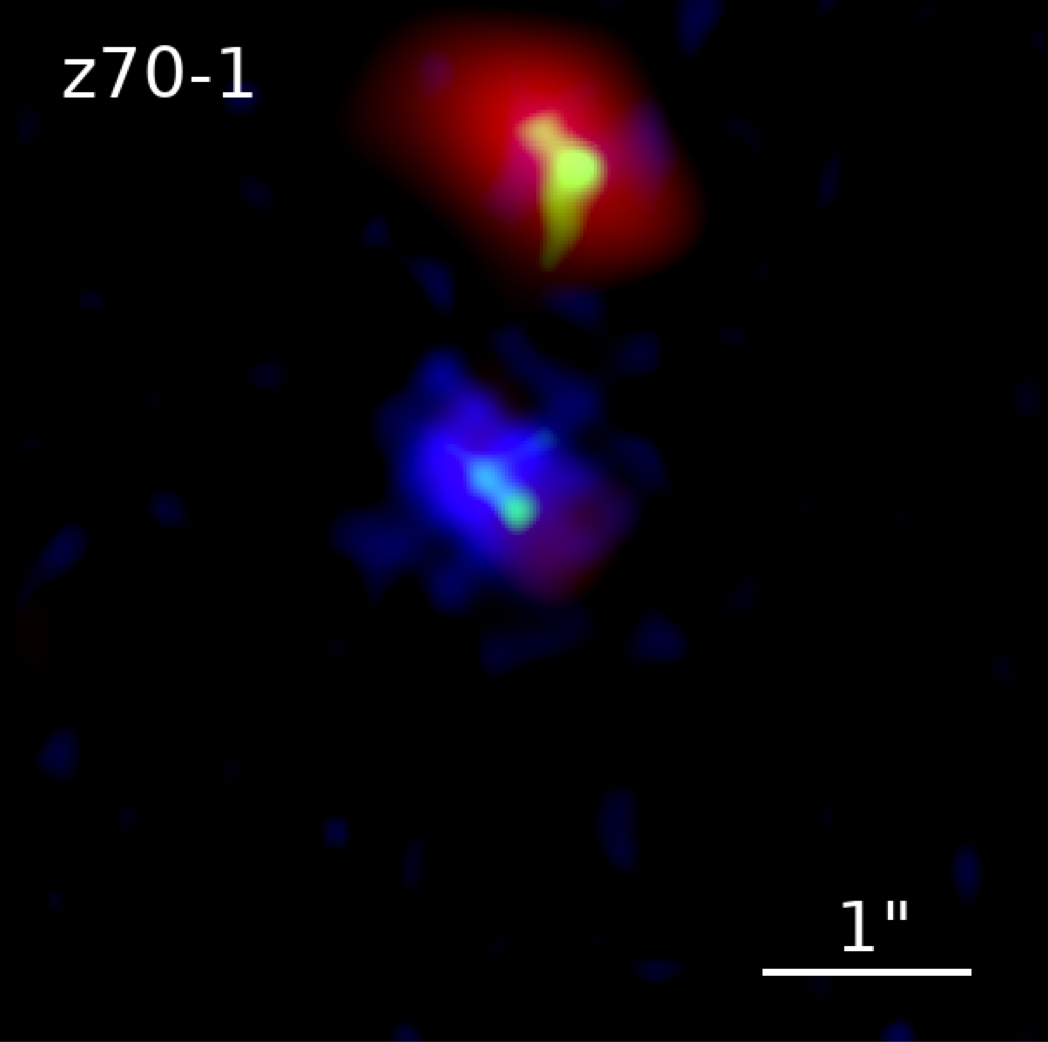}{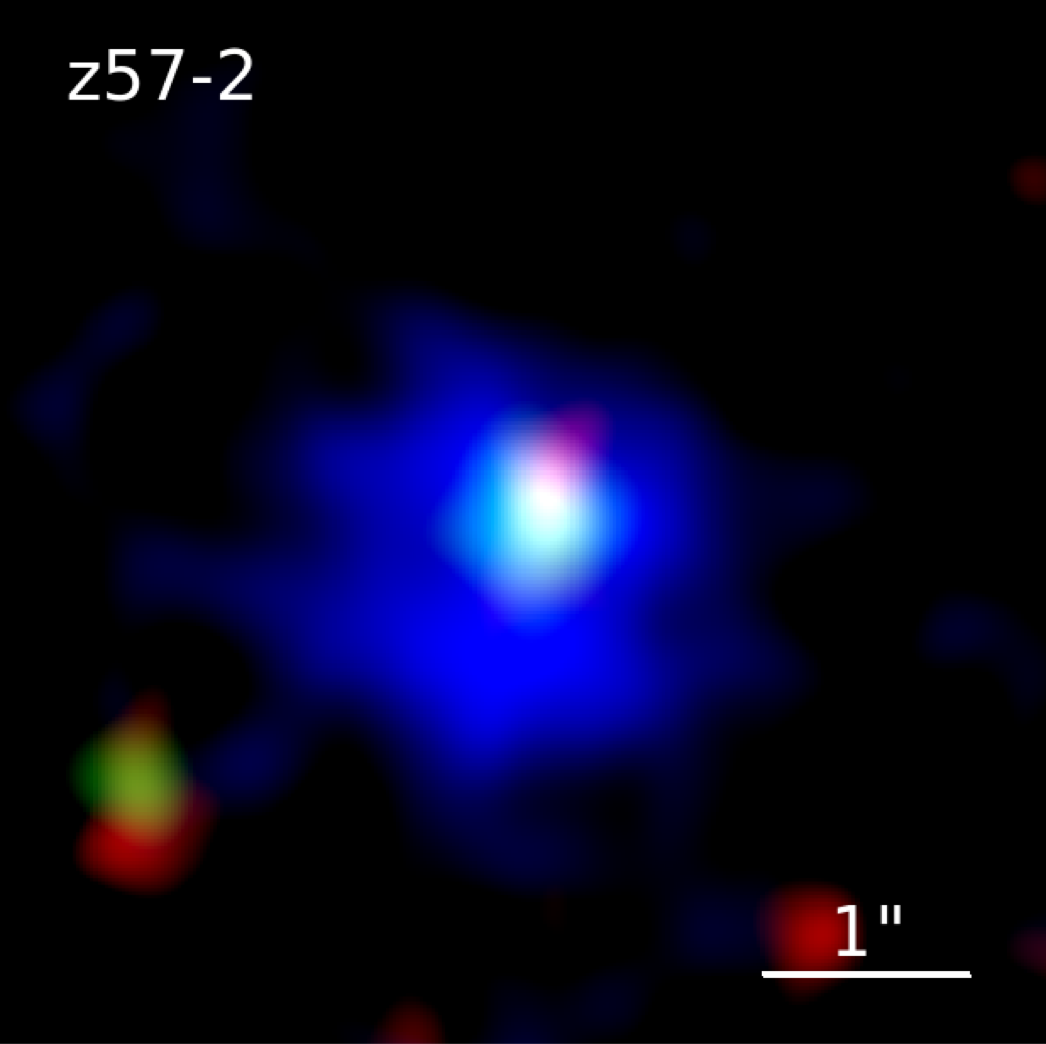}
    \caption{Composite pseudocolor images of z70-1 (left) and z57-2 (right). The upper object in the left panel is a foreground source. The RGB colors of z70-1 are presented with 3.6 $\mu$m, $y$, and NB973 images, respectively. For z57-2, the RGB colors correspond to $y$, $z$, and NB816 images, respectively. Because z57-2 does not show a clear center in the NB816 image, we smooth the $y$, $z$, and NB816 images of z57-2 with a Gaussian kernel whose sigma value is 0\farcs17 before we make the pseudocolor image. The size of the images is $5\arcsec\times5\arcsec$. The length of $1\arcsec$ is indicated as a white bar.}
    \label{fig:RGB}
\end{figure}

\begin{figure}[ht]
    \centering
    \plotone{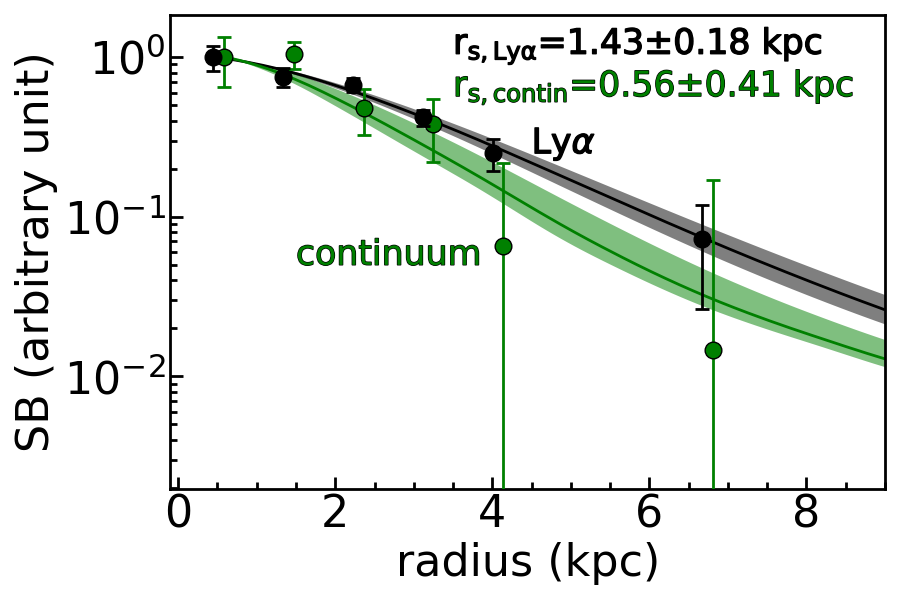}
    \caption{Ly$\alpha$ (black) and continuum (green) surface brightness profiles of z70-1. The filled circles are the profiles measured from images. The solid lines present the best-fit exponential functions. The shaded regions indicate the errors of the best-fit functions. We normalize all the profiles at the radius of $\sim$0.5 kpc for comparison. To avoid overlaps, the continuum profile is slightly shifted along the horizontal axis by +0.15.}
    \label{fig:LAB70_radial}
\end{figure}

In this study, we have identified the most distant LAB found to date, z70-1 at $z=7.0$. The composite pseudocolor image of z70-1 is presented in Figure \ref{fig:RGB} left. Figure \ref{fig:LAB70_radial} shows the Ly$\alpha$ and continuum profiles of z70-1. To test whether the Ly$\alpha$ profile of z70-1 is more extended than the continuum profile, we fit the exponential function shown in Equation \ref{eq:4} to the Ly$\alpha$ and continuum profiles. In the fitting, the errors of the profiles are considered. The best-fit scale lengths of the Ly$\alpha$ and continuum profiles are $1.43\pm0.18$ and $0.56\pm0.41$ kpc, respectively. We estimate the statistical significance of the difference between the scale lengths of Ly$\alpha$ and continuum profiles assuming a normal distribution. We find that the Ly$\alpha$ and continuum profiles are different at the 87\% confidence level. This suggests that the Ly$\alpha$ emission of z70-1 is more extended than the continuum. Taken together with the identification of the Ly$\alpha$ emission line on the spectrum, and the bright Ly$\alpha$ luminosity of z70-1, our result suggests that z70-1 is a real LAB at $z=7.0$.

\subsection{An Extremely Diffuse LAB at $z=5.7$}

\begin{figure}[ht]
\epsscale{1.16}
    \centering
    \plotone{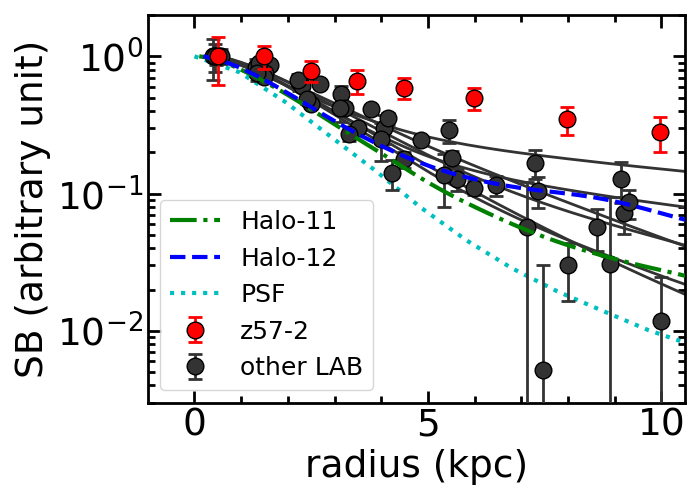}
    \caption{Ly$\alpha$ surface brightness profiles of z57-2 (red filled circles), the other 6 LABs (black filled circles), and 2 model galaxies of Halo-11 (green dash-dotted line) and Halo-12 (blue dashed line). The black solid lines are the best-fit total models of the other 6 LABs in Figure \ref{fig:fit}. The cyan dotted line represents the PSF. The profiles of Halo-11 and Halo-12 are convolved with the PSF. All the profiles are normalized at the radius of $\sim$0 kpc for comparison.}
    \label{fig:yajima_model}
\end{figure}

The NB816 image of z57-2 in Figure \ref{fig:4} suggests that z57-2 has very diffuse Ly$\alpha$ emission presenting no clear center, which is apparently different from the other 6 LABs. The composite pseudocolor image of z57-2 is shown in Figure \ref{fig:RGB} right. Figure \ref{fig:yajima_model} displays the Ly$\alpha$ surface brightness profile of z57-2, together with the other 6 LABs and 2 model galaxies of Halo-11 and Halo-12 (\citealt{yajima2017, arata2018}) at $z\sim 6$. Cosmological hydrodynamic and radiative transfer simulations produce Halo-11 and Halo-12 that have halo masses of $1.6\times10^{11}$ and $7.5\times10^{11}\ M_{\odot}$, respectively. As suggested by \citet{behroozi2013}, the halo masses of Halo-11 and Halo-12 correspond to stellar masses of $\sim2.0\times10^{9}$ and $1.4\times10^{10}\ M_{\odot}$ at $z=6.0$, respectively, which are consistent with the stellar masses of our LABs estimated by the SED fitting (Section 4.4). In Figure \ref{fig:yajima_model}, it is clear that z57-2 has a more diffuse Ly$\alpha$ profile than the other 6 LABs. Moreover, model galaxies of Halo-11 and Halo-12 cannot explain the extremely diffuse Ly$\alpha$ profile of z57-2.

The spectrum in Figure \ref{fig:spec-lya} shows that z57-2 has a Ly$\alpha$ emission line with a FWHM of $\sim600$ km/s that is broader than those of the other 6 LABs. It should be also noted that the Ly$\alpha$ line of z57-2 shows multiple peaks. These features may be caused by dynamical systems, such as multiple components or mergers. Another possibility is that z57-2 has a nearly static cloud of thick {\sc Hi} gas that resonantly scatters Ly$\alpha$ photons produced at the center of this system. The static cloud should have varying {\sc Hi} column densities that cause the positional dependence 
of the Ly$\alpha$ line center and line width found in Figure \ref{fig:lc_fwhm}.

\subsection{Physical Origin of the diffuse Ly$\alpha$ emission of LABs}
Previous studies have suggested several physical origins of the diffuse Ly$\alpha$ emission around a galaxy, including scenarios of photoionization, Ly$\alpha$ resonant scattering, cooling radiation, outflows, and satellite galaxies. We discuss these scenarios separately below.

\citet{masribas2016} suggest that fluorescence can generate Ly$\alpha$ photons that account for diffuse Ly$\alpha$ emission around LAEs at $z=3.1$. As for LABs, fluorescence is very likely to happen because of the large abundance of ionizing photons expected from the bright $M_{\rm UV}$. \citet{geach2009} and \citet{overzier2013} argue that fluorescence alone can explain the luminous and diffuse Ly$\alpha$ emission of LABs hosting AGNs. Because our LABs have bright $M_{\rm UV}$ and possible AGN activities, fluorescence may be the origin of the diffuse Ly$\alpha$ emission.

In the scenario of Ly$\alpha$ resonant scattering, the Ly$\alpha$ photons are emitted by star formation in the galaxy center. Although our LABs have very high Ly$\alpha$ luminosities, the Ly$\alpha$ equivalent widths are $\sim50-200$ $\rm \AA$, consistent with the Ly$\alpha$ equivalent width from dust-free star formation estimated in \citet{charlot1993}. This consistency suggests that the diffuse Ly$\alpha$ emission can be explained by resonantly scattered Ly$\alpha$ photons generated in the star-forming galaxy center.

Gravitational cooling radiation may also play an important role in generating an extended Ly$\alpha$ emission. Using Ly$\alpha$ radiative transfer models of LAEs with a mean stellar mass of $2.9\times10^{10}$ $M_{\odot}$ at $z=3.1$, \citet{lake2015} show that cooling radiation can contribute 40-55\% of the total Ly$\alpha$ luminosity within a virial radius of 56 kpc. On the other hand, if cooling radiation is the major origin of the diffuse Ly$\alpha$ emission, the Ly$\alpha$ equivalent width would likely be greater than 240 $\rm \AA$ that is the maximum equivalent width predicted by stellar models \citep{charlot1993}. Although the Ly$\alpha$ equivalent widths of our LABs are smaller than 240 $\rm \AA$, it should be noted that the Ly$\alpha$ escape fractions of our LABs might be low. For example, z49-1 has a Ly$\alpha$ escape fraction of 0.11 as we discussed in Section 4.4. Because the low Ly$\alpha$ equivalent widths of our LABs may be caused by low Ly$\alpha$ escape fractions, we cannot rule out the possibility that cooling radiation is the origin of the diffuse Ly$\alpha$ emission.

Using an analytical model and a high-resolution hydrodynamic simulation, respectively, \citet{taniguchi2000} and \citet{mori2004} suggest that outflows driven by multiple supernova explosions are able to produce extended Ly$\alpha$ emission with a Ly$\alpha$ luminosity of $\sim 10^{43}$ erg s$^{-1}$. This Ly$\alpha$ luminosity is consistent with those of our LABs. On the other hand, our LABs may have starbursts driven by possible mergers as suggested by the multiple UV components in Figure \ref{fig:4}. Multiple supernova explosions are likely to happen in starbursts, and drive outflows that produce the luminous and diffuse Ly$\alpha$ emission of our LABs.

In Figure \ref{fig:4}, the HST images of z49-1, z66-1, and z66-2 clearly show multiple UV continuum components. It is likely that having multiple UV continuum components is a common feature of high-$z$ LABs. The multiple components may correspond to multiple star forming clumps in one galaxy, mergers, or satellite galaxies. It is possible that satellite galaxies are responsible for the large continuum size of LABs. On the other hand, if the satellite galaxies are the major contributors to the diffuse Ly$\alpha$ emission, one would expect that the Ly$\alpha$ and continuum profiles have the similar shape even if the satellite galaxies are not resolved. In Figure \ref{fig:fit}, it should be noted that the core component has the same shape as the continuum profile, and that the Ly$\alpha$ profile cannot be explained by the single core component. However, the difference between the Ly$\alpha$ and continuum profiles can be caused by satellite galaxies with high Ly$\alpha$ equivalent widths, such as the faint LAEs at $z=2.9-6.7$ found in \citet{maseda2018}. It is possible that satellite galaxies with high Ly$\alpha$ equivalent widths are the origin of the diffuse Ly$\alpha$ emission around LABs.

In conclusion, all of the 5 scenarios of fluorescence, resonant scattering, gravitational cooling radiation, outflows, and satellite galaxies may contribute to the diffuse Ly$\alpha$ emission around LABs.

\section{Summary}
In this study, we investigate the photometric and spectroscopic properties of seven LABs; two LABs at $z=4.888$ (z49-1) and $z=6.965$ (z70-1) identified by us, and five previously-known LABs at $z=5.7-6.6$ (z57-1, z57-2, z66-1, z66-2, and z66-3). Our results are summarized below.

\begin{enumerate}
    \item We find that z70-1 has extended Ly$\alpha$ emission with a scale length of $1.4\pm0.2$ kpc that is about three times larger than the UV continuum. The object of z70-1 is the most distant LAB identified to date.
    
    \item We show that z57-2 has Ly$\alpha$ emission that is much more diffuse than the other 6 LABs. The origin of the extremely diffuse Ly$\alpha$ emission of z57-2 is unclear, and cannot be explained by cosmological hydrodynamic and radiative transfer simulations.
    
    \item We measure the core and halo scale lengths of the Ly$\alpha$ profiles of our LABs, and show that our LABs are consistent with the extrapolation of the relations between scale lengths and galaxy properties ($L_{\mathrm{Ly\alpha}}$, EW$_{0}$, and M$_{\mathrm{UV}}$) of MUSE LAEs from \citet{leclercq2017}. This suggests that our LABs and MUSE LAEs have similar connections between the diffuse Ly$\alpha$ emission and properties of the host galaxies, and that typical LABs at $z\gtrsim5$ are not special objects, but star-forming galaxies at the bright end.
    
    \item We investigate the large scale structure around our LABs by measuring the LAE overdensity. We find that all the 7 LABs except z66-2 are located in overdensed regions. Our LABs show no significant correlation between the halo scale length and LAE overdensity.
    
    \item The 7 LABs except z49-1 exhibit no AGN signatures such as X-ray emission, {\sc Nv}$\lambda$1240, or Ly$\alpha$ line broadening. The object of z49-1 has a strong {\sc Civ}$\lambda$1548 emission line that suggests an AGN. We compare the {\sc Civ} equivalent width and {\sc Civ}/He{\sc ii} ratio of z49-1 with the AGN and SFG models in \citet{nakajima2018}, and find that z49-1 is an AGN candidate although the possibility of a young and low-metallicity SFG cannot be eliminated. 
    
    \item We find that all the Ly$\alpha$ emission lines of the 7 LABs show velocity gradients on the spectra. The Ly$\alpha$ velocity gradients and line widths of z49-1 and z57-2 are larger than those of the other 5 LABs, which may be caused by an AGN (not likely for z57-2), mergers, or dense neutral hydrogen gas in the {\sc Hi} region.
    
    \item We discuss the physical origin of the diffuse Ly$\alpha$ emission around our LABs. Fluorescence, resonant scattering, gravitational cooling radiation, outflows, and satellite galaxies can contribute to the diffuse Ly$\alpha$ emission.

\end{enumerate}

The Hyper Suprime-Cam (HSC) collaboration includes the astronomical communities of Japan and Taiwan, and Princeton University. The HSC instrumentation and software were developed by the National Astronomical Observatory of Japan (NAOJ), the Kavli Institute for the Physics and Mathematics of the Universe (Kavli IPMU), the University of Tokyo, the High Energy Accelerator Research Organization (KEK), the Academia Sinica Institute for Astronomy and Astrophysics in Taiwan (ASIAA), and Princeton University. Funding was contributed by the FIRST program from Japanese Cabinet Office, the Ministry of Education, Culture, Sports, Science and Technology (MEXT), the Japan Society for the Promotion of Science (JSPS), Japan Science and Technology Agency (JST), the Toray Science Foundation, NAOJ, Kavli IPMU, KEK, ASIAA, and Princeton University. 

This paper makes use of software developed for the Large Synoptic Survey Telescope. We thank the LSST Project for making their code available as free software at  http://dm.lsst.org

The Pan-STARRS1 Surveys (PS1) have been made possible through contributions of the Institute for Astronomy, the University of Hawaii, the Pan-STARRS Project Office, the Max-Planck Society and its participating institutes, the Max Planck Institute for Astronomy, Heidelberg and the Max Planck Institute for Extraterrestrial Physics, Garching, The Johns Hopkins University, Durham University, the University of Edinburgh, Queen’s University Belfast, the Harvard-Smithsonian Center for Astrophysics, the Las Cumbres Observatory Global Telescope Network Incorporated, the National Central University of Taiwan, the Space Telescope Science Institute, the National Aeronautics and Space Administration under Grant No. NNX08AR22G issued through the Planetary Science Division of the NASA Science Mission Directorate, the National Science Foundation under Grant No. AST-1238877, the University of Maryland, and Eotvos Lorand University (ELTE) and the Los Alamos National Laboratory.

Based in part on data collected at the Subaru Telescope and retrieved from the HSC data archive system, which is operated by Subaru Telescope and Astronomy Data Center at National Astronomical Observatory of Japan.

The NB718 and NB816 filters were supported by Ehime University (PI: Y. Taniguchi). The NB921 and NB973 filters were supported by KAKENHI (23244025) Grant-in-Aid for Scientific Research (A) through the Japan Society for the Promotion of Science (PI: M. Ouchi). 

Some of the data presented herein were obtained at the W. M. Keck Observatory, which is operated as a scientific partnership among the California Institute of Technology, the University of California and the National Aeronautics and Space Administration. The Observatory was made possible by the generous financial support of the W. M. Keck Foundation.

The authors wish to recognize and acknowledge the very significant cultural role and reverence that the summit of Maunakea has always had within the indigenous Hawaiian community.  We are most fortunate to have the opportunity to conduct observations from this mountain.

This paper includes data gathered with the 6.5 meter Magellan Telescopes located at Las Campanas Observatory, Chile.

This work is supported by World Premier International Research Center Initiative (WPI Initiative), MEXT, Japan, and KAKENHI (15H02064, 17H01110, and 17H01114) Grant-in-Aid for Scientific Research (A) through Japan Society for the Promotion of Science.

\end{document}